\documentclass[11pt]{article}
\pdfoutput=1

\usepackage{a4wide}
\usepackage{Packages}

\usepackage{Definitions}

\newcommand{\OfficialTitle}{Sublattice Counting and Orbifolds}

\author{
  \begin{minipage}{.97\linewidth}
    \vspace{1cm}
    \begin{center}
      \begin{small}
        \textbf{Amihay Hanany}$^{1}$, \textbf{Domenico Orlando}$^{2}$ and \textbf{Susanne Reffert}$^{2}$
     \end{small}
    \end{center}
    \vspace{1cm} \hspace{2cm}\begin{minipage}{.7\linewidth}
      {\it \begin{footnotesize}
          \begin{itemize}
          \item[${}^1$] Theoretical Physics Group, The Blackett Laboratory\\
            Imperial College London, Prince Consort Road \\
            London, SW7 2AZ, UK
          \item[${}^2$] Institute for the Mathematics and Physics of
              the Universe, \\The University of Tokyo, Kashiwa-no-Ha
              5-1-5, \\ Kashiwa-shi, 277-8568 Chiba, Japan. \\
          \end{itemize}
        \end{footnotesize}}
    \end{minipage}
    \vspace{1cm}
  \end{minipage}
}

\date{}

\title{\vspace{3cm}
  \begin{huge}
    \textbf{\OfficialTitle}
  \end{huge}
}

\begin{document}

\numberwithin{equation}{section}

\begin{titlepage}
  \maketitle
  \thispagestyle{empty}

  \vspace{-14cm}
  \begin{flushright}
    Imperial/TP/10/AH/01 \\
    IPMU10-0025
  \end{flushright}

  \vspace{14cm}

  \begin{center}
    \textsc{Abstract}\\
  \end{center}

  Abelian orbifolds of $\setC^3$ are known to be encoded by hexagonal
  brane tilings. To date it is not known how to count all such
  orbifolds. We fill this gap by employing number theoretic techniques
  from crystallography, and by making use of Polya's Enumeration
  Theorem. The results turn out to be beautifully encoded in terms of
  partition functions and Dirichlet series. The same methods apply to
  counting orbifolds of any toric non--compact Calabi--Yau
  singularity. As additional examples, we count the orbifolds of the
  conifold, of the $L^{aba}$ theories, and of $\setC^4$.

\end{titlepage}

\setstretch{1.1}

\tableofcontents

\section{Introduction}\label{sec:intro}

\emph{Brane tilings}~\cite{Hanany:2005ve,Franco:2005rj} have met with
a lot of interest in the past few years. Each brane tiling gives rise
to a \emph{quiver gauge theory}, which can describe either the theory
living on D3 branes probing a toric Calabi--Yau--three singularity, or
the theory living on M2 branes probing a toric Calabi--Yau--four
singularity~\cite{Hanany:2008cd}. Faces, edges and nodes of the brane
tiling -- a periodic bipartite tiling of the plane -- correspond
respectively to gauge groups, chiral bifundamental fields and
interaction terms in the superpotential. A periodic quiver can be
constructed from the brane tiling by substituting nodes by faces and
edges by arrows. The faces of the periodic quiver thus represent terms
in the superpotential with a (negative) positive sign for (anti--)
clockwise orientation and are in fact extensions of the usual notion
of a quiver which does not have a ``built--in'' superpotential.

Recently, the question of enumerating \emph{all} possible brane
tilings has been raised~\cite{Davey:2009bp}. A classification of
all brane tilings with up to and including $N_T=6$ terms in the
superpotential was given in this paper, and the results for all brane
tilings with $N_T=8$ terms were computed but not published. These
results were derived using a computer code that reaches its limits for
$N_T=10$ terms. Thus a need for a better algorithm or a different approach
arises. One possible approach is to count the number of tilings for a
fixed number of terms $N_T$ and collect the answer into a generating
function. This turns out to be a difficult task and to date an answer
is still unknown.  We can therefore simplify the problem further and
attempt to count the number of ``sub-tilings''.  For example, we
can ask how many tilings with $n$ hexagons in the fundamental domain
there are. Or we can ask how many inequivalent tilings with $2n$
squares one can construct.  Such questions turn out to be relatively
easy to solve and are the subject of the present work.

It is known that all orbifolds of a given geometry correspond to a
repetition of the fundamental domain~\cite{Hanany:2005ve}. For example, the number
of inequivalent hexagonal tilings with $n$ tiles equals
the number of inequivalent orbifolds of $\setC^3$ with an
Abelian group of order $n$. Similarly, the number of tilings with $2n$
squares is the number of inequivalent orbifolds of the conifold by an
Abelian group of order $n$, etc.  Note that in the case of \emph{compact} Calabi--Yau
manifolds, the problem is different because of the gluing conditions for the
patches that result in a finite number of admissible orbifolds.

It is useful to map the problem of counting brane tilings to the
problem of counting sub-lattices.  Take for example the problem of
counting the Abelian orbifolds of $\setC^3$. Instead of counting the
lattices obtained by the repetition of $n$ tiles, we can take the
standard bipartite hexagonal lattice and count its sublattices of index $n$.
Similarly, the problem of counting Abelian
orbifolds of the conifold is equivalent to counting a certain type of square
sublattices.  We are thus led to the subject of \emph{enumeration of
sublattices} which has been studied by the crystallography community in
great detail (see
\emph{e.g.}~\cite{Kucab:1981,Rutherford:1904p2035,Baake:1997a,Rutherford:2009p1973}
and references therein). Fortunately, we are able to take results from
this field and apply them to the questions of interest in this paper.

In the following we will outline some methods for counting the
sublattices of a given lattice. While it is possible to enumerate by
hand the first few orbifolds, this quickly becomes cumbersome. A
general understanding of how the number of sublattices $f(n)$ with
\emph{index} (\ie size of the fundamental cell) $n$ behaves and grows
is therefore desirable. It turns out that this number decomposes into
the symmetries of a given lattice. Our main results are closed
formulae for the number of Abelian orbifolds of $\setC^3$, of the
conifold, of $L^{aba}$ theories and of $\setC^4$, which are
furthermore generalizable to any toric non--compact Calabi--Yau. In
all the cases corresponding to Calabi--Yau--three geometries we find
that for large $n$ the dominant contribution is
\begin{equation}
  f(n) \sim \frac{\sigma(n)}{\abs{G}}  \, , \hspace{2em} \text{for $n \gg 1$ , }   
\end{equation}
where $G$ is the symmetry group of the fundamental cell of the brane
tiling, and $\sigma$ is the sum--of--divisors function.

The plan of this note is as follows. In
Section~\ref{sec:what-we-are-counting}, we outline the problem we are
studying. In the following sections, we introduce the knowledge
necessary to count all the sublattices of a given lattice and show how
to apply it to the case of the hexagonal lattice (Abelian orbifolds of
$\setC^3$). In Section~\ref{sec:symm}, we discuss the \emph{cycle
  index} which captures the symmetries of a given lattice. The
\emph{Hermite normal form} is introduced in
Section~\ref{sec:hermite-normal-form}. In Section~\ref{sec:gen}, the
concepts of the \emph{Dirichlet convolution} and the \emph{Dirichlet
  series} are introduced.  In Section~\ref{sec:examples} we discuss
the examples of the square lattice (Abelian orbifolds of the
conifold), the $L^{aba}$ theories, and the tetrahedral lattice
(Abelian orbifolds of $\setC^4$) following the same steps as detailed
in the previous sections for the case of the hexagonal lattice. In
Appendix~\ref{sec:generic-lattice-d}, we briefly discuss the general
lattice in $d$ dimensions.

\section{What we are counting and how}
\label{sec:what-we-are-counting}

To get a better understanding of our counting problem, let us give a
brief description of some details. A more exhaustive explanation will
appear in a forthcoming publication~\cite{Rak} which will describe
several computer codes that were used in order to obtain some of the
results used in this note.

We start by looking at orbifolds of $\setC^3$. For simplicity,
let us focus on $\setZ_n$ orbifolds. More general Abelian
orbifolds such as $\setZ_n\times\setZ_m$ can be treated in a
similar fashion. Let us denote the coordinates of $\setC^3$ by
$\set{z_1, z_2, z_3}$, and the orbifold action by $(a_1, a_2, a_3)$ such
that $\set{z_1, z_2, z_3 }\sim \set{ \omega^{a_1} z_1,\omega^{a_2} z_2,
\omega^{a_3} z_3}$, with $\omega^n=1$ and $a_1+a_2+a_3=0 \mod n$.  In
this notation, the problem is to find all triples $(a_1, a_2, a_3)$
that give inequivalent orbifolds of $\setC^3$. The first few
cases are as follows. For $n=1$, the orbifold group is trivial and
there is only one case, $\setC^3$. For $n=2$, there is again one
case, which is commonly denoted in the literature as
$\setC^2/\setZ_2\times \setC$. For $n=3$, there are two
cases, $\setC^2/\setZ_3\times \setC$ with orbifold
action $(1,2,0)$ and $\setC^3/\setZ_3$ with orbifold
action $(1,1,1)$. For $n=4$, there are three cases,
$\setC^2/\setZ_4\times \setC$,
$\setC^3/\setZ_4$ with orbifold action $(1,1,2)$ and
$\setC^3/\setZ_2\times \setZ_2$, etc. The brane tilings
for the first examples can be found in Table~\ref{tab:C3-tilings}; a
count of the first 16 cases can be found in the last row,
$f^{\triangle}$, of Table~\ref{tab:triangle-numbers}.

\newcolumntype{S}{>{\centering\arraybackslash} m{.17\linewidth} }

\begin{table}
  \centering
  \begin{tabular}{SSSSS}
    \toprule
    $n$ & $ 1$ & $2$ & $3$ & $3$ \\ \midrule
    brane tiling & \includegraphics[width=.17\textwidth]{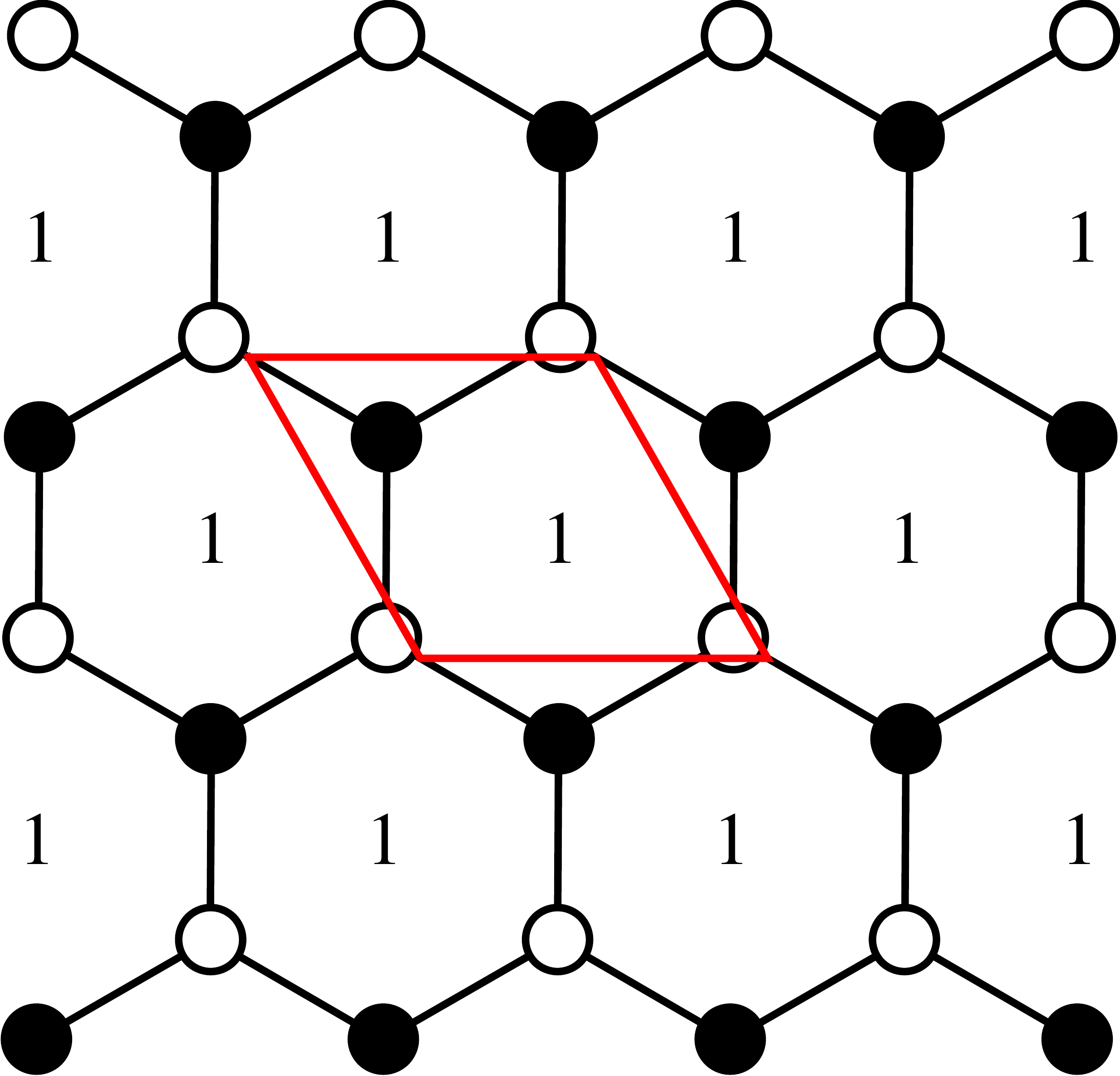} & \includegraphics[width=.17\textwidth]{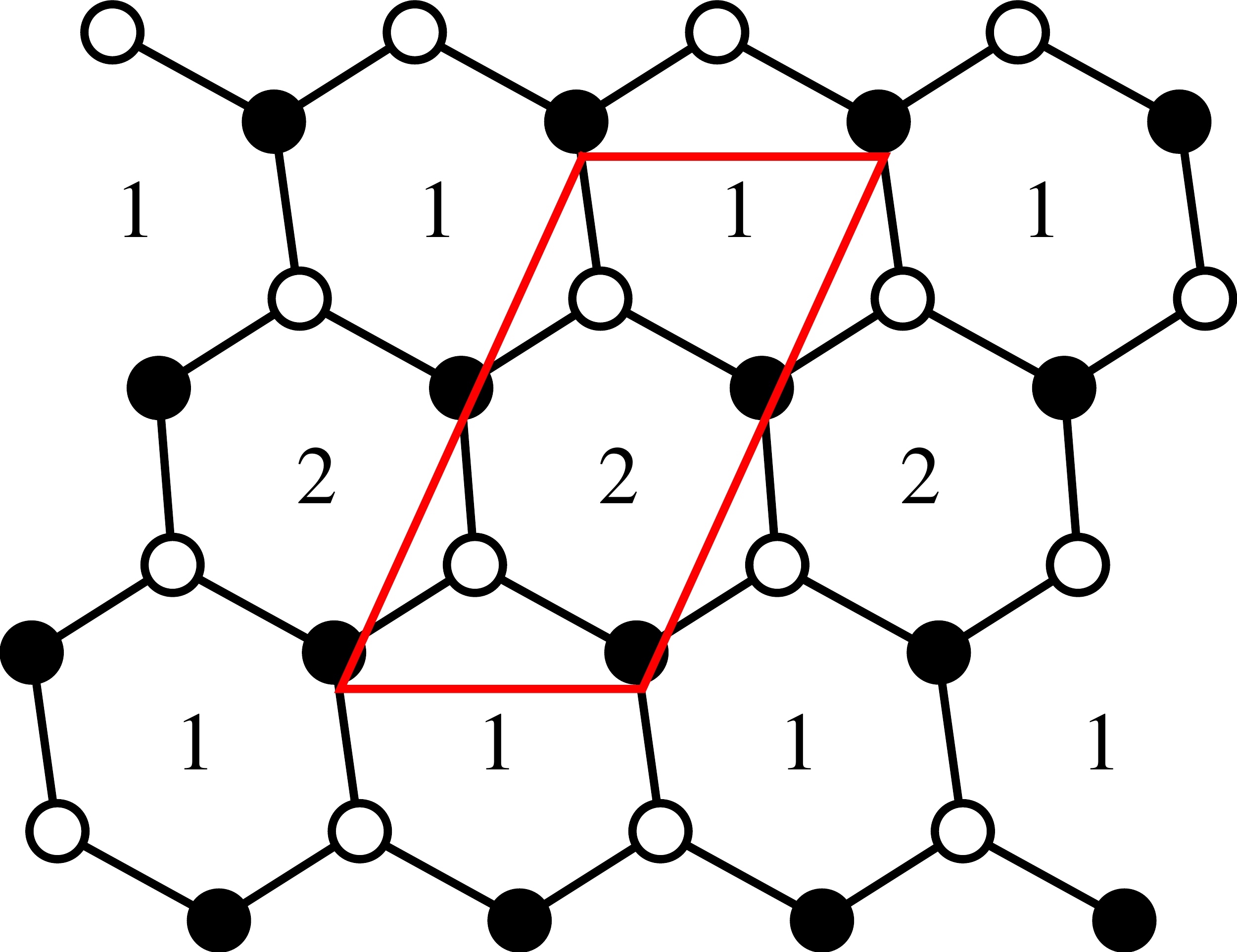} & \includegraphics[width=.17\textwidth]{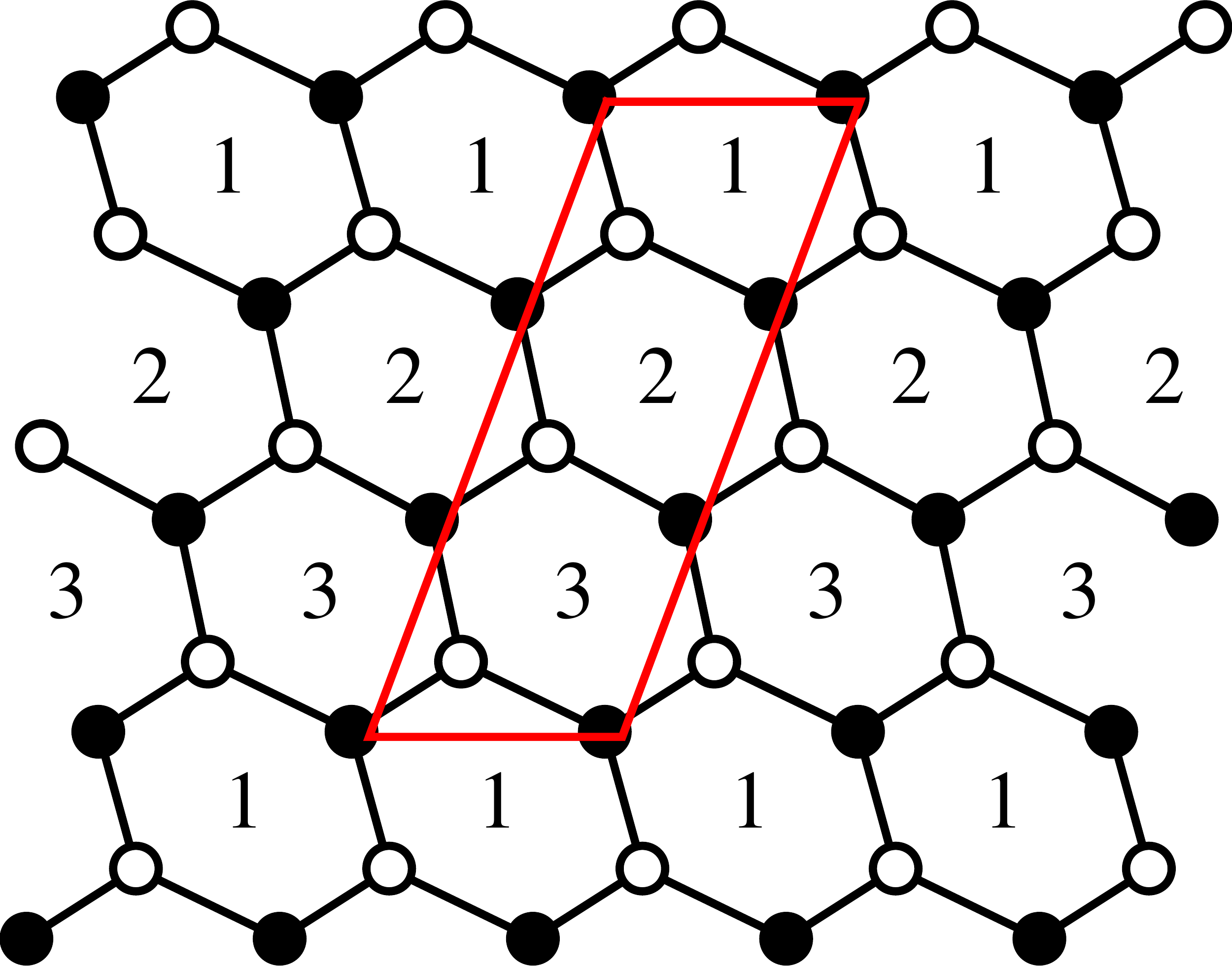}  & \includegraphics[width=.17\textwidth]{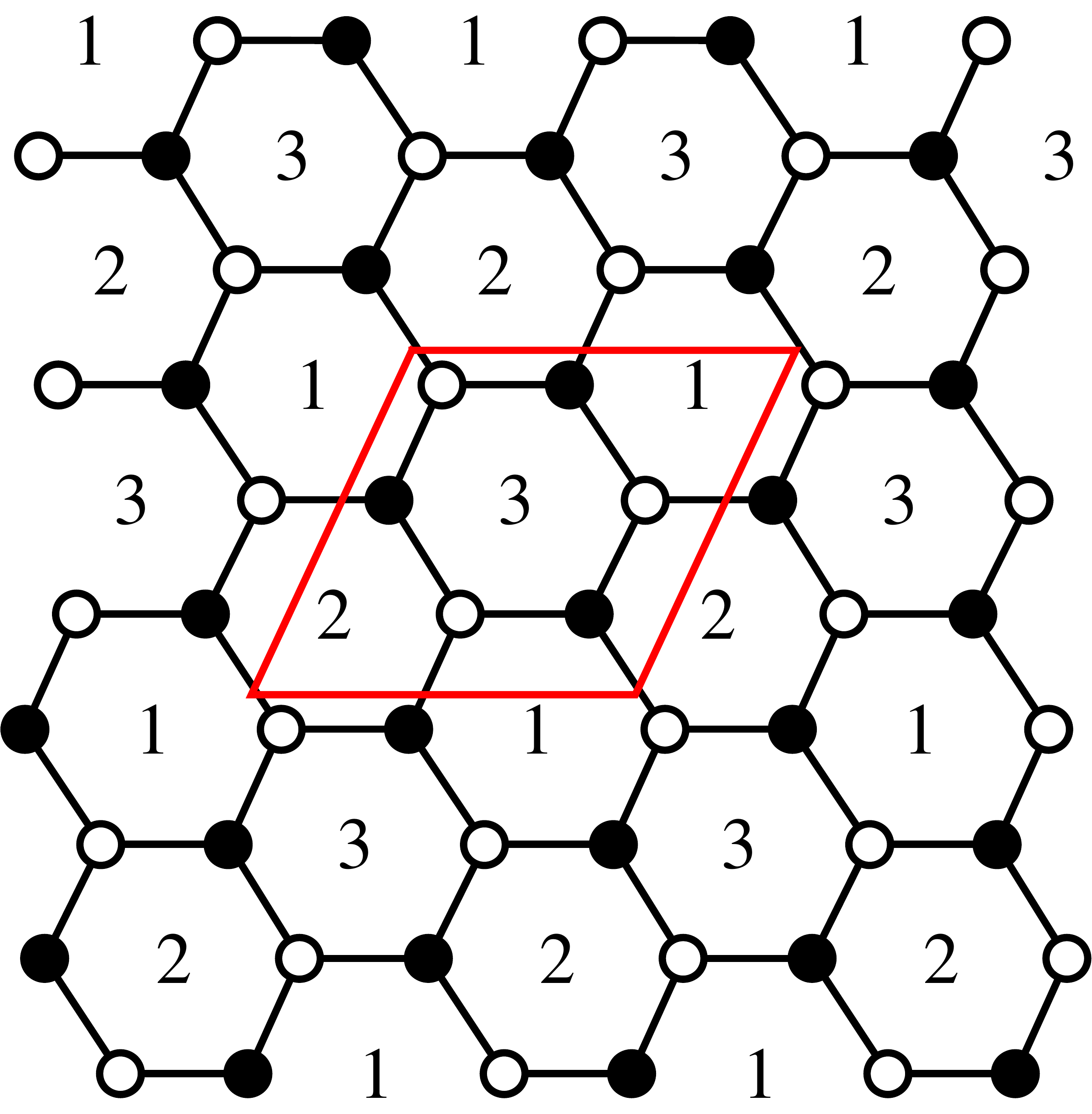}  \\
    geometry & $ \setC^3 $ & $\setC^2 / \setZ_2 \times \setC$ & $\setC^2 / \setZ_3 \times \setC$ & $ \setC^3/\setZ_3$ \\ \bottomrule
  \end{tabular}
  \caption{Brane tilings for the first orbifolds of $\setC^3$.}
  \label{tab:C3-tilings}
\end{table}

An alternative way of formulating the problem is by looking at the
\emph{toric diagrams} of these orbifolds. Since the toric diagram of
$\setC^3$ is a triangle of unit area, the toric diagram of an
orbifold of $\setC^3$ by an Abelian group of order $n$ is
again a triangle but with an area which is $n$ times larger. The
problem of counting all inequivalent orbifolds of $\setC^3$ is
therefore equivalent to the problem of finding all triangles with vertices
on integral points and area $n$. Of course, since these are toric
diagrams, two triangles which are related by a $GL(2, \mathbb {Z})$
transformation are equivalent. This provides another approach to
counting orbifolds. A systematic method of finding such triangles is
algorithmically different from that of finding inequivalent orbifold
actions and therefore provides an alternative approach to the counting
problem. As stated above, more details can be found
in~\cite{Rak}.

Yet a third approach is to think of the brane tiling as forming a
bipartite hexagonal lattice and the problem of finding inequivalent
toric diagrams is mapped to the problem of finding its
sublattices. This is going be the language used in the next
sections. The first step is to identify the symmetries of each
lattice, which is the topic of the next section.

There is however a subtle point that needs to be emphasized. The brane
tiling encodes the same information as the quiver diagram plus the
superpotential, and in certain cases, one toric geometry admits
several possible quiver gauge theories. In this case we speak of
different \emph{toric phases}. The lattices of the brane tilings pertaining
to the different toric phases are different, but  they
preserve the same symmetries, and for the enumeration of toric geometries,
the symmetries are what matters.  Even though our construction works
starting from any lattice (brane tiling), if we want to count all the
resulting quiver gauge theories, we need to keep track of the
different toric phases that can appear in the process of orbifolding a
given geometry.  Consider for example the case of the conifold (see
Section~\ref{sec:square}). The brane tiling is a bipartite square
lattice. Dividing by $\setZ_2$, one obtains $\mathbb{F}_0$. This
variety has two different toric phases, one corresponds to a
sublattice of the square lattice, the other one is represented
by a square--octagon lattice. While the quiver gauge theory of the
former lattice is captured by counting the orbifolds of the conifold,
the gauge theories stemming from the latter have to be considered
separately.  The final consequence is that even if we use the lattice
of the brane tiling (which corresponds directly to the gauge theory),
our counting covers the geometries but in general \emph{not} all the
gauge theories or toric phases that can arise from the orbifolds of a
given geometry.

\section{Symmetries and the cycle index}\label{sec:symm}

In the following, we need a way to capture the symmetries of a given
lattice. Let us label the vertices of the fundamental cell by the
numbers $\set{1, \dots, m }$. We now want to describe the group of
permutations $G$ of the set $X=\set{1, \dots, m}$ which result in the
same fundamental cell. The \emph{cycle index} encodes this
information~\cite{Biggs:2003}.  For the admissible symmetries, the
fact that we are considering bipartite lattices plays an important
role, since the preservation of the coloring of the vertices results
in an additional constraint.

In our case, it is most convenient to express the permutations in
\emph{cycle notation}. The cycles of $g\in G$ are the orbits of
the elements $\varepsilon\in X$ under $g$. For each group element $g$
we start with $\varepsilon_1\in X$ and write down its orbit in
parentheses, $(\varepsilon_1 \, g(\varepsilon_1) \, g^2(\varepsilon_1)
\, \dots \, g^{k-1}(\varepsilon_1))$, where $g^{k}(\varepsilon_1) =
\varepsilon_1$.  We continue to do the same with the next element that
has not yet appeared in an orbit until we have exhausted all the
elements of $X$. Each $g \in G$ can thus be expressed in terms of
$\alpha_k $ disjoint cycles of length $k$; cycles of length one
correspond to elements that are fixed under $g$.  The \emph{type} of
$g$ is given by the partition of $m$ $[1^{\alpha_1} 2^{\alpha_2} \dots
l^{\alpha_l}]$, where $m = \alpha_1 + 2 \alpha_2 + ... + l \alpha_l $.
The partition is represented by the expression
\begin{equation}
  \zeta_g( x_1, \dots, x_l ) = x_1^{\alpha_1} x_2^{\alpha_2} \dots x_l^{\alpha_l} \, . 
\end{equation}
\begin{definition}
  The \emph{cycle index} of $G$ is obtained by summing the $\zeta_g$ over all
  elements $g\in G$ and dividing by the number of elements $|G|$:
  \begin{equation}
    \label{eq:cycle-index}
    Z_G (x_1, \dots , x_l ) = \frac{1}{\abs{G}} \sum_{g \in G} \zeta_g( x_1, \dots, x_l ) = \frac{1}{\abs{G}} \sum_{\alpha} c(\alpha_1,...,\alpha_l)\, x_1^{\alpha_1}\cdots x_l^{\alpha_l} \, ,
  \end{equation}
  where $c(\alpha_1,...,\alpha_l)$ is the number of permutations of
  type $[1^{\alpha_1} 2^{\alpha_2} \dots l^{\alpha_l}]$, and the sum
  runs over all partitions of $m$.
\end{definition}

\begin{table}
  \centering
  \begin{tabular}{clccccl}
    \toprule 
    &$g$ & $\alpha_1$& $\alpha_2$& $\alpha_3$& $c(\alpha_i) $ & $\zeta$ \\ 
    \midrule
    \begin{minipage}{.1\linewidth}
      \vspace{1em}
        \includegraphics[scale=1]{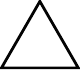}
        \begin{picture}(20,0)(,) 
          \begin{footnotesize}
            \put(-5,5){$1$} \put(25,5){$2$} \put(10,37){$3$}
          \end{footnotesize}
      \end{picture}
    \end{minipage} & $1$ & $3$ & - & - & $1$ & $x_1^3$ \\  
    \begin{minipage}{.1\linewidth}
      \vspace{1em}
        \includegraphics[scale=1]{triangle}
        \begin{picture}(20,0)(,) 
          \begin{footnotesize}
            \put(-5,5){$2$} \put(25,5){$1$} \put(10,37){$3$}
          \end{footnotesize}
      \end{picture}
    \end{minipage} & $(12)$ & $1$ & $1$ & - & $3$ & $x_1 x_2$ \\  
    \begin{minipage}{.1\linewidth}
      \vspace{1em}
        \includegraphics[scale=1]{triangle}
        \begin{picture}(20,0)(,) 
          \begin{footnotesize}
            \put(-5,5){$3$} \put(25,5){$1$} \put(10,37){$2$}
          \end{footnotesize}
      \end{picture}
    \end{minipage} & $(123)$ & - & - & $1$ & $2$ & $x_3$ \\  
    \bottomrule
  \end{tabular}
  \caption{Cycle index for the symmetric group $S_3$. $Z_{S_3} = \frac{1}{6} \left( x_1^3 + 3 x_1 x_2 + 2 x_3 \right)$.}
  \label{tab:S3-cycle-index}
\end{table}

\bigskip

Some cases of interest are:
\begin{enumerate}
\item The \emph{cyclic group} $C_n$ is the group of symmetries
  associated to a circular object where reflections are excluded:
  \begin{equation}
    Z(C_m) = \frac{1}{m} \sum_{d|m} \varphi(d) x_d^{m/d} \, ,
  \end{equation}
  where $\varphi(d)$ is the \emph{totient function}.
\item The \emph{dihedral group} $D_m$ is the group of symmetries
  associated to a circular object where reflections are allowed:
  \begin{equation}
    Z(D_m) = \frac{1}{2} Z(C_m) +
    \begin{cases}
      \frac{1}{2} x_1 x_2^{(m-1)/2} \, , &\text{if $m$ is odd, } \\
      \frac{1}{4} \left( x_1^2 x_2^{(m-2)/2} + x_2^{m/2} \right) \, ,
      & \text{ if $m$ is even.}
    \end{cases}  
  \end{equation}
\item The \emph{symmetric group} $S_m$ is the group of all permutations of $n$ symbols:
  \begin{equation}
    Z(S_m) = \sum_{\alpha_1+2 \alpha_2 + \dots + k j_k = m} \frac{1}{\prod_{k=1}^m k^{\alpha_k} \alpha_k! } \prod_{k=1}^m x_k^{\alpha_k} \, .
  \end{equation}
A convenient recursion formula is given by
\begin{equation}
    Z(S_m) = \frac{1}{m}\sum_{k=1}^m x_k Z(S_{m-k}), \qquad Z(S_{0})=1 .
  \end{equation}
\end{enumerate}

The link between the cycle index and the number of sublattices of the
lattice $L$ is provided by Burnside's lemma.
\begin{lemma}
  Let $G$ be a group of permutations of the set $X$. The number $N(G)$
  of \emph{orbits} of $G$ is given by the average over $G$ of the sizes
  of the fixed sets:
  \begin{equation}
    N(G) = \frac{1}{\abs{G}} \sum_{g \in G} \abs{F_g} \, ; \hspace{2em} F_g = \set{ x \in X | g(x) = x } \, .
  \end{equation}
\end{lemma}
The number $f^L(n)$ of sublattices of index $n$ can be understood as
the number of orbits of the symmetry group $G$ when acting on the set
$X_n$ of sublattices of index $n$. According to the lemma, this can be
written as the average of the number of elements in $X_n$ that are
left invariant by the action of $g \in G$:
\begin{equation}
  f^L(G) = \frac{1}{\abs{G}} \sum_{g \in G} f^L_{g} (n) \, ; \hspace{2em} f^L_g (n) = \abs{ \set{ x \in X_n | g(x) = x } } \, .  
\end{equation}
Using the cycle decomposition introduced above we can rewrite this
expression as a sum over the types of the elements $g$, indexed by
partitions $\alpha $:
\begin{equation}
 \label{eq:sequence-decomposition}
  f^L(n) = \frac{1}{\abs{G}} \sum_\alpha c(\alpha) f^L_{\mathbf{x}^{\alpha}}(n) \, .
\end{equation}
By comparison with Equation~\eqref{eq:cycle-index} we see that we obtain a
subsequence for each monomial in the cycle index $Z_G(x)$.

\bigskip

Every group contains the identity, which is represented by the
partition $[1^m]$. The corresponding number of invariant sublattices
$f^L_{x_1^m}(n)$ only depends on the dimension of the lattice $d =
\dim [L]$ and is given by the formula in
Equation~(\ref{eq:generic-lattice-power-series}):
\begin{equation}
  \label{eq:generic-lattice-f(n)}
  f^L_{x_1^m} (n) = \sum_{\substack{k_0, \dots, k_{d-1} = 1 \\ k_0 k_1 \cdots k_{d-1} = n}}^n k_1 k_2^2 \cdots k_{d-1}^{d-1} \, .  
\end{equation}

\begin{figure}
  \centering
  \includegraphics[width=.8\textwidth]{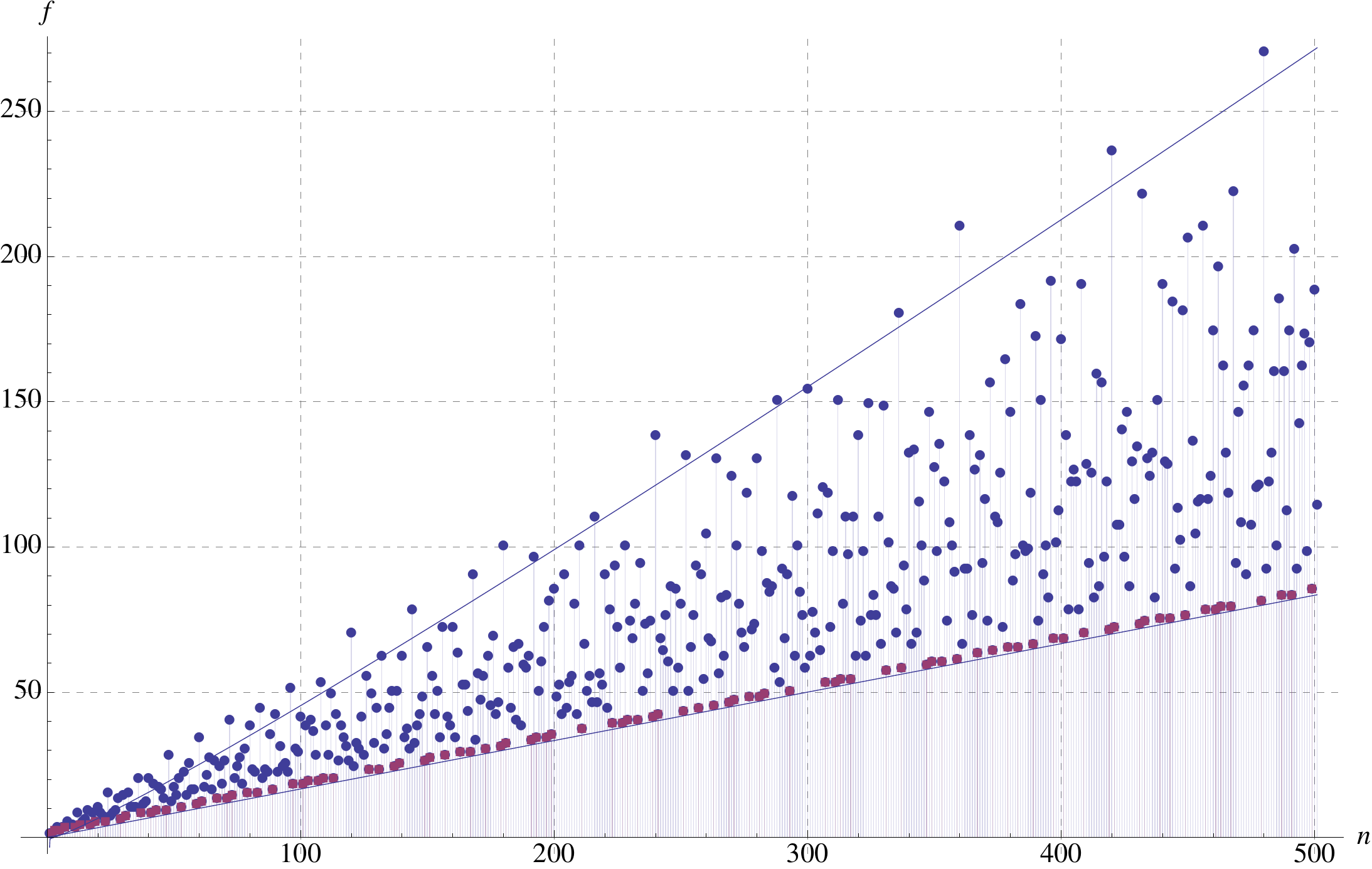}
  \caption{Scatter plot of the sequence $f^\triangle$ for a hexagonal
    lattice. Prime numbers are emphasized in red. The two lines correspond to
    $n/6 $ and $e^\gamma n \log \log n / 6 $.}
  \label{fig:triangle-scatter}
\end{figure}

\begin{example}
  Consider the bipartite hexagonal lattice corresponding to the
  geometry of $\setC^3$. Because of the bipartiteness, the
  symmetry group is not $D_6$ as expected for a hexagon, but $S_3$,
  which we will also denote by a
  triangle. Table~\ref{tab:S3-cycle-index} shows the cycle
  decomposition for the symmetric group
  $S_3$:
  \begin{equation}
    Z_{S_3} = \frac{1}{6} \left( x_1^3 + 3\, x_1 x_2 + 2\, x_3 \right) \, .
  \end{equation}
  Table~\ref{tab:triangle-numbers} gives the number of sublattices of
  index $n$ for the bipartite hexagonal lattice for each of the
  monomials appearing in the expression above.
  \begin{equation}
    f^{\triangle} = \frac{1}{6} \left( f^{\triangle}_{x_1^3} + 3\, f^{\triangle}_{x_1 x_2} + 2\, f^{\triangle}_{x_3} \right)
  \end{equation}
  The numbers
  $f^\triangle (n) $, $ n \le 500$ are represented in
  Figure~\ref{fig:triangle-scatter}, where the prime numbers are
  emphasized.
\end{example}

\begin{table}
  \centering
  \begin{tabular}{ccccccccccccccccccccc}
    \toprule 
    $n$ & 1 & 2 & 3 & 4 & 5 & 6 & 7 & 8 & 9 & 10 & 11 & 12 & 13 & 14 & 15 & 16 \\
    \midrule
    $f^{\triangle}_{x_1^3}$ & 1 & 3 & 4 & 7 & 6 & 12 & 8 & 15 & 13 & 18 & 12 & 28 & 14 & 24 & 24 & 31 \\
    $f^{\triangle}_{x_1 x_2}$ & 1 & 1 & 2 & 3 & 2 & 2 & 2 & 5 & 3 & 2 & 2 & 6 & 2 & 2 & 4 & 7 \\
    $f^{\triangle}_{x_3}$ & 1 & 0 & 1 & 1 & 0 & 0 & 2 & 0 & 1 & 0 & 0 & 1 & 2 & 0 & 0 & 1 \\
    \midrule
    $f^{\triangle}$ & 1 & 1 & 2 & 3 & 2 & 3 & 3 & 5 & 4 & 4 & 3 & 8 & 4 & 5 & 6 & 9 \\
    \bottomrule
  \end{tabular}
  \caption{Number of sublattices of index $n$ for the hexagonal lattice, classified by the cycles of the symmetric group $S_3$. According to the cycle index decomposition, $f^{\triangle} = \frac{1}{6} \left( f^{\triangle}_{x_1^3} + 3 f^{\triangle}_{x_1 x_2} + 2 f^{\triangle}_{x_3} \right)$.}
  \label{tab:triangle-numbers}
\end{table}

\section{Hermite normal form}
\label{sec:hermite-normal-form}

Consider a lattice $L_d$ generated by the $d$ vectors
$\braket{y_1, \dots, y_d}$. Any sublattice $L^\prime$ of $L_d$ is
generated by $d$ vectors $\braket{x_1, \dots, x_d}$ which can be
written as
\begin{equation}\label{eq:Hermite}
  \begin{cases}
    x_1 = a_{11} y_1 \\
    x_2 = a_{21} y_1 + a_{22} y_2 \\
    \dots \\
    x_d = a_{d1} y_1 + a_{d2} y_2 + \dots + a_{dd} y_d\,,
  \end{cases}
\end{equation}
where the integer coefficients $a_{ij}$ satisfy the conditions
\begin{equation}\label{eq:cond1}
  0 \le a_{ij} < a_{ii} \hspace{2em} \forall j < i \, .
\end{equation}
The above construction is the so--called \emph{Hermite normal form}.
The \emph{index} $n$ of the lattice is given by the product
\begin{equation}\label{eq:cond2}
  n = \prod_{i=1}^d a_{ii} \, .  
\end{equation}
Expressing the sublattices via Equation~(\ref{eq:Hermite}), 
the problem of counting the sublattices of a generic lattice turns into the
problem of counting the number of matrices $a_{ij}$ that satisfy the
conditions Equation~(\ref{eq:cond1}) and Equation~(\ref{eq:cond2}).

\bigskip

Consider a two--dimensional lattice $L_2$. The condition $a_{11}
a_{22} = n $ can be satisfied by the choice $a_{22} = m $ and $a_{11}
= n/m$, where $m$ is a divisor of $n$. If we want to count the number
of sublattices invariant under the symmetry $\mathbf{x}^\alpha $, we
need to enumerate the possible values of $a_{21}$. The constraint
$a_{21} < a_{22}$ introduces a dependence of the number of
possible values of $a_{21}$, $\#\set{ a_{21} } =
g^{L_2}_{\mathbf{x}^\alpha } (a_{22} )$, on $a_{22}$. The total number of
sublattices $f^{L_2}_{\mathbf{x}^\alpha } (n)$ is thus given by summing $g^{L_2}_{\mathbf{x}^\alpha } (m )$ over all
the divisors of $n$:
\begin{equation}
\label{eq:fL2-as-convolution}
  f^{L_2}_{\mathbf{x}^\alpha } (n) = \sum_{m|n} g^{L_2}_{\mathbf{x}^\alpha} (m ) \, .  
\end{equation}
Repeating the same construction for a three--dimensional lattice $L_3$, we find that
\begin{equation}
  \label{eq:fL3-as-convolution}
  f^{L_3}_{\mathbf{x}^\alpha } (n) = \sum_{m_1|n} \sum_{m_2|m_1} h^{L_3}_{\mathbf{x}^\alpha} (\frac{m_1}{m_2} ) l^{L_3}_{\mathbf{x}^\alpha } ( m_2 ) \, . 
\end{equation}
A similar decomposition into $d -1 $ nested sums appears for any
lattice of dimension $d$.

\section{Generating functions}
\label{sec:gen}

\subsection{The Dirichlet convolution}
\label{sec:dirichl-conv}

Looking at Figure~\ref{fig:triangle-scatter}, one realizes that
prime numbers play a special role. This is one of the clues that point to the fact that
the sequences corresponding to the monomials of the cycle index have
the property of being \emph{multiplicative}~\cite{Apostol:1998}.  A
possible explanation for this may be linked to the observation that the
orbifold group $\setZ_{p q}$ is isomorphic to the group
$\setZ_{p} \times \setZ_{q}$ for $p, q$ primes.
\begin{definition}
  A sequence $f$ is multiplicative if
  \begin{equation}
    f( n m ) = f(n) f(m) \, , \quad \text{when $(n,m) = 1$} \, ,
  \end{equation}
  where $(n,m)$ denotes the \emph{greatest common divisor} between
  $n$ and $m$.
\end{definition}
It follows in particular that $f$ is completely determined by its
values for primes and their
powers, since for any $n$ we can use the factorization $ n = p_1^{a_1}
p_2^{a_2} \dots p_r^{a_r} $ , and
\begin{equation}
  f (n) = f(p_1^{a_1}) f( p_2^{a_2}) \dots f(p_r^{a_r}) \, . 
\end{equation}

Multiplicative sequences form a group under the \emph{Dirichlet convolution}. For
our examples, it is convenient to use this property and decompose each
of the sequences into products of other sequences that are easier
to deal with. First, we need the following
\begin{definition}
  The \emph{Dirichlet convolution} of two sequences $f$ and $g$ is the
  sequence $h$ defined by
  \begin{equation}
    f(n) = ( g * h ) (n) = \sum_{m|n} g(m)\, h ( \frac{n}{m} ) \, ,
  \end{equation}
  where the notation $m|n$ means that the sum runs over all the divisors $m$ of $n$.
\end{definition}
One can prove that this convolution is commutative, $f*g  = g*f$, and
associative, $f * \left( g* h \right) = \left( f * g \right) * h$, and
that the sequence $\id $ defined by
\begin{equation}
  \id (n) = \set{ 1, 0, 0, \dots } \, 
\end{equation}
is the identity, $f * \id = f$. To each sequence $f$ one can associate
its inverse  $f^{-1}$ satisfying
\begin{equation}
  f * f^{-1} = f^{-1} * f = \id \, .  
\end{equation}
The inverse can be evaluated recursively via
\begin{equation}
  f^{-1}(n) = - \frac{1}{f(1)} \sum_{\substack{d|n\\d < n}} f (\frac{n}{d} ) f^{-1}(d) \, .  
\end{equation}

We have observed above that in two dimensions, the number of invariant
sublattices can be put in the form of
Equation~\eqref{eq:fL2-as-convolution},
\begin{equation}
  f^{L_2}_{\mathbf{x}^\alpha } (n) = \sum_{m|n} g^{L_2}_{\mathbf{x}^\alpha} (m ) \, .  
\end{equation}
This is equivalent to the statement that the sequence
$f^{L_2}_{\mathbf{x}^\alpha }$ is the convolution of
$g^{L_2}_{\mathbf{x}^{\alpha }}$ with the unit $\un$ defined by
\begin{equation}
  \un (n) = \set{ 1, 1, 1, \dots } \, .
\end{equation}
Its inverse is the \emph{Möbius function} defined by
\begin{equation}
  \mu(n) = \begin{cases}
    1 & \text{if $n$ is square--free with an even number of distinct prime factors,} \\
    -1 & \text{if $n$ is square--free with an odd number of distinct prime factors,} \\
    0 & \text{otherwise.}
  \end{cases}
\end{equation}
It follows that if $f = g * \un $, then $g = \mu * f $.

\begin{example}
  In the case of the bipartite hexagonal lattice we find:
  \begin{enumerate}
  \item The sequence $f^{\triangle}_{x_1^3} = \set{1, 3, 4, 7, 6, 12,
      8, 15, \dots }$ corresponds to the identity
    permutation $x_1^3$ and it is given by Equation~(\ref{eq:generic-lattice-f(n)})
    with $d=2$. It can also be written as the convolution
    \begin{equation}
      f^{\triangle}_{x_1^3} = \un * \N \, ,
    \end{equation}
    where
    \begin{equation}
      \N(n) = \set{1, 2, 3, \dots } \, .
    \end{equation}
  \item The sequence $f^{\triangle}_{x_1 x_2} = \set{1, 1, 2, 3, 2, 2,
      2, 5, \dots} $ can be written as the convolution of a periodic
    sequence of period $4$ and the unit:
    \begin{equation}
      f^{\triangle}_{x_1 x_2} = \set{1, 0, 1, 2, 1, 0, 1, 2, 1, \dots } * \un \, .
    \end{equation}
    $g^\triangle_{x_1 x_2} $ is in turn the convolution of a finite
    sequence and $\un$:
    \begin{equation}
      f^{\triangle}_{x_1 x_2} = \set{1, 0, 1, 2, 1, 0, 1, 2, 1, \dots } * \un =
      \set{1, -1, 0, 2} * \un * \un \, . 
    \end{equation}
  \item The last sequence $f^{\triangle}_{x_3} = \set{1, 0, 1, 1, 0,
      0, 2, 0, \dots }$ also has the form of the convolution of the
    unity with a periodic sequence of period $3$:
    \begin{equation}
      f^{\triangle}_{x_3} = \set{ 1, -1, 0, 1, -1, 0, 1, -1, 0, \dots } * \un \, .
    \end{equation}
    The periodic sequence is the (non--principal) \emph{Dirichlet character}
    of modulus three (see~\cite{Apostol:1998}):
    \begin{equation}
      g^\triangle_{x_3} = \chi_{3,2} (n) = \set{ 1, -1, 0, 1, -1, 0,  \dots } \, .
    \end{equation}
  \end{enumerate}
  Putting all together we find that the sequence $f^{\triangle}$ can
  be written as
  \begin{equation}
    f^{\triangle} = \frac{1}{6} \left( f^{\triangle}_{x_1^3} + 3\, f^{\triangle}_{x_1 x_2} + 2\,
    f^{\triangle}_{x_3}  \right) = \frac{1}{6} \left(  \N + 3 \set{1,0,-1,2} * \un +
      2\, \chi_{3,2} \right)
    * \un \, .
  \end{equation}
\end{example}

\subsection{Dirichlet series and power series}
\label{sec:dirichlet-series}

The information contained in a sequence $f$ can be usefully encoded
into a \emph{generating function} (more commonly used as a \emph{partition function} in the physics literature). In the following, we will use two types of generating functions:
\begin{enumerate}
\item the formal power series (partition function)
  \begin{equation}
   F(t) = \sum_{n=1}^\infty f(n) t^n \, ; 
  \end{equation}
\item the Dirichlet series
  \begin{equation}
    F(s) = \sum_{n=1}^\infty \frac{f(n)}{n^s} \, .   
  \end{equation}
\end{enumerate}
The corresponding inverse transformations are given by
\begin{gather}
  f(n) = \frac{1}{2\pi \imath} \oint \frac{F(t)}{t^{n+1}} \, \di t \, ,\\
  f(n) = \lim_{T\to \infty} \frac{1}{2T} \int_{-T}^T \left. F( s ) n^s \right|_{s = \sigma +
  \imath \tau}  \, \di \tau \, .
\end{gather}

Dirichlet series are appropriate in the case of multiplicative
sequences. In particular, if $f$ is multiplicative, the series can be
expanded in terms of an infinite product over the primes, the
\emph{Euler product}:
\begin{equation}
  F(s) = \sum_{n=1}^\infty \frac{f(n)}{n^s} = \prod_{p} \left( 1 +
    \frac{f(p)}{p^s} + \frac{f (p^2)}{p^{2s}} + \dots  \right) \, .
\end{equation}
This is consistent with the observation that a multiplicative
sequence is determined by the values taken for powers of prime
numbers.

\bigskip

Let us now consider the sequences that appeared in the example above:
\begin{enumerate}
\item For the identity $\id$:
  \begin{align}
    \id (s) = 1 \, , && \id (t) = t \, ;
  \end{align}
\item For the unit $\un$ we obtain Riemann's zeta function:
  \begin{align}
    \un (s) = \sum_{n=1}^\infty \frac{1}{n^s} = \zeta(s) \, , && \un (t) = \frac{1}{1-t} - 1 \, ;
  \end{align}
\item For $\N$:
  \begin{align}
    \N (s) = \sum_{n=1}^\infty \frac{n}{n^s} = \zeta(s -1 ) \, , && \N (t) = \frac{1+t^3}{\left( 1 - t \right ) \left( 1- t^2 \right)} - 1 ;
  \end{align}
\item For the finite sequence $\set{1, -1, 0 ,2 } $:
  \begin{align}
    \set{1, -1, 0 ,2 } (s) = 1 - \frac{1}{2^s} + \frac{2}{4^s} \, , && \set{1,-1,0,2} (t) = t -t^2 +2 t^4 \, ;
  \end{align}
\item For the Dirichlet character $\chi_{3,2}$, the corresponding Dirichlet series
  is the so--called $L$--\emph{function}
  \begin{align}
    \chi_{3,2} (s) = \sum_{n=1}^\infty \frac{\chi_{3,2}(n)}{n^s} = L(s, \chi_{3,2}
    ) \, , && \chi_{3,2} (t) = \frac{\left( 1 + t \right) \left( 1 - t^2 \right)}{1-t^3} - 1 \, .
  \end{align}
\end{enumerate}

Both types of generating functions have a simple behavior under
Dirichlet convolution. Let $f, g $ and $h$ be such that
\begin{equation}
 f = g * h \, . 
\end{equation}
The power series for $h$ reads:
\begin{equation}
 F (t) = \sum_{n=1}^\infty f(n) n^n = \sum_{n=1}^\infty \sum_{m|n}
 g(m)\, h(\frac{n}{m})\, t^n = \sum_{k=1}^\infty \sum_{m=1}^\infty g(m)\,
 h(k)\, t^{m k } \, . 
\end{equation}
This can be expressed in two ways, using the generating function for
$g$ or for $h$:
\begin{equation}
 F(t) = \sum_{m=1}^\infty g(m) H(t^m) = \sum_{k=1}^\infty h(k) G(t^k)
 \, . 
\end{equation}
In particular, since all our sequences can be written as sums over
divisors (or equivalently as Dirichlet convolutions with the unit), we
will always write
\begin{equation}
  F(t) = \sum_{k=1}^\infty G(t^k) \, .  
\end{equation}
It is also possible to write the power series for the inverse of the
Dirichlet convolution as follows. Let
\begin{equation}
  f(t) = \sum_{k,m=1}^\infty g(m)\, h(k)\, t^{m k} \, ,
\end{equation}
then 
\begin{equation}
  H(t) = \sum_{k=1}^\infty h(k)\, t^k = \sum_{m=1}^\infty \mu(k)\, g(k)\, F(t^k) \, ,
\end{equation}
where $\mu $ is the Möbius function.

\bigskip 

More directly, the Dirichlet series is decomposed as
\begin{equation}
  F(s) = \sum_{n=1}^\infty \frac{f(n)}{n^s} = \sum_{n=1}^\infty \sum_{m|n} \frac{g(m)\, h(\frac{n}{m})}{n^s} = \sum_{k=1}^\infty \sum_{m=1}^\infty \frac{g(m)\, h(k)}{m^s k^s} = G(s) H(s) \, .
\end{equation}
Note that the Dirichlet series corresponding to a sequence can be
also understood as the \emph{Laplace transform} of a discrete
measure. This explains why it exchanges convolution and pointwise
products.

Both generating functions can be seen as linear transformations, hence
the decomposition in Equation~\eqref{eq:sequence-decomposition} still holds.
\begin{example}
  Let us now apply these formulas to the terms in $f^\triangle$:
  \begin{enumerate}
  \item The sequence $f^{\triangle}_{x_1^3}$ is decomposed as
    $f^{\triangle}_{x_1^3} = \un * \N $, hence the corresponding power
    series is generated by
    \begin{equation}
      G^\triangle_{x_1^3} (t) = \sum_{k=1}^\infty k\, t^{k} = \frac{1+t^3}{\left( 1  - t  \right) \left( 1 - t^2 \right)} - 1 \, ,    
    \end{equation}
    and the Dirichlet series reads
    \begin{equation}
      F^\triangle_{x_1^3} (s) = \zeta(s) \zeta(s-1) \, .
    \end{equation}
  \item The sequence $f^\triangle_{x_1 x_2} = \set{1,-1,0,2} * \un * \un $
    gives
    \begin{gather}
      G^\triangle_{x_1 x_2}(t) = \frac{1+t^3}{\left( 1 - t \right) \left( 1 + t^2 \right)} - 1 \, , \\
      F^\triangle_{x_1 x_2} (s) = \left( 1 - 2^{-s} + 2^{1-2s} \right)
      \zeta(s)^2 \, .
    \end{gather}
  \item The sequence $f^\triangle_{x_3} = \chi_{3,2} * \un $ gives
    \begin{gather}
      G^\triangle_{x_3} (t) = \frac{\left( 1 + t \right) \left( 1 - t^2 \right)}{1 - t^3} - 1 \, ,\\
      F^\triangle_{x_3} (s) = L(s,\chi_{3,2}) \zeta(s) \, .
    \end{gather}
  \end{enumerate}

  Collecting all the terms we find:
  \begin{enumerate}
  \item For the power series:
    \begin{equation}
      G^\triangle(t) = \frac{1}{6}\, G^\triangle_{x_1^3} (t) + \frac{1}{2}\, G^\triangle_{x_1 x_2}(t) + \frac{1}{3}\, G^\triangle_{x_3} (t) = \frac{1}{\left( 1 - t \right) \left( 1 + t^2 \right) \left( 1 - t^3 \right)} - 1 \, ,
    \end{equation}
    whence
    \begin{equation}
      \boxed{F^\triangle(t) = \sum_{m=1}^\infty \left[ \frac{1}{\left( 1 - t^m \right) \left( 1 + t^{2m} \right) \left( 1 - t^{3m} \right)} - 1 \right] = \sum_{m=1}^\infty \sum_{\substack{n_1, n_2, n_3 = 0 \\ \neq \left( 0, 0, 0 \right)}}^\infty \left( -  \right)^{n_2} t^{m \left( n_1 + 2 n_2 + 3 n_3  \right)} }\, .
    \end{equation}
  \item For the Dirichlet series:
    \begin{equation}
      \label{eq:Dirichlet-triangle}
      \boxed{F^\triangle(s) = \frac{\zeta(s)}{6} \left( \zeta(s-1) + 3 \left( 1 - 2^{-s} + 2^{1-2s} \right) \zeta(s) + 2 L(s,\chi_{3,2}) \right) } \, .
    \end{equation}
  \end{enumerate}
  The generating functions for the hexagonal lattice are summarized in
  Table~\ref{tab:Dirichlet-triangle}.
\end{example}

\begin{table}
  \centering
  \begin{tabular}{cccc}
    \toprule
    symmetry  & Dirichlet series $G(s) $ & power series $G(t)$ \\
    \midrule
    $x_1^3$  & $\zeta(s-1) $ & $\displaystyle{\frac{1+t^3}{\left( 1  - t  \right) \left( 1 - t^2 \right)} - 1}$ \\
    $x_1 x_2 $ & $\left( 1 - 2^{-s} + 2^{1-2s} \right) \zeta(s) $ & $\displaystyle{\frac{1+t^3}{\left( 1 - t \right) \left( 1 + t^2 \right)} - 1} $ &\\
    $x_3 $ & $L(s,\chi_{3,2}) $ & $\displaystyle{\frac{\left( 1 + t \right) \left( 1 - t^2 \right)}{1 - t^3} - 1}$ \\ 
    \bottomrule
  \end{tabular}
  \caption{Generating functions for the sublattices of the bipartite hexagonal lattice organized by symmetry. The actual sequences correspond to $F(s) = G(s) \zeta(s) $ and $F(t) = \sum_{k=1}^\infty G(t^k)$. }
  \label{tab:Dirichlet-triangle}
\end{table}

\subsection{Asymptotic behavior}
\label{sec:asymptotic-behavior}

The asymptotic behavior of a sequence can be derived by looking at the
corresponding Dirichlet series. 
\begin{theorem}
  Let $F(s)$ be a Dirichlet series with non--negative coefficients
  that converges for $\Re(s) > \alpha > 0$, and suppose that $F(s)$ is
  holomorphic in all points of the line $\Re(s) = \alpha $, except for
  $s = \alpha $. If for $s \to \alpha^+$, the Dirichlet series behaves
  as
  \begin{equation}
    F(s) \sim A(s) + \frac{B(s)}{\left(s - \alpha  \right)^{m+1}} \, ,
  \end{equation}
  where $m \in \setN$, and both $A(s) $ and $B(s) $ are holomorphic in
  $s = \alpha $, then the partial sum of the coefficients is
  asymptotic to:
  \begin{equation}
    \sum_{n = 1 }^N a_n \sim \frac{B(\alpha )}{\alpha \, m!}
    N^\alpha \log^m (N) \, .
  \end{equation}
\end{theorem}
In order to apply this theorem, we can make use of the following facts:
\begin{enumerate}
\item The Riemann zeta function $\zeta(s)$ is analytic everywhere,
  except for a simple pole at $s=1$ with residue $1$;
\item The $L$--function $L(s,\chi)$ is analytic everywhere, except for
  a simple pole at $s=1$ if $\chi $ is a principal character.
\end{enumerate}

\bigskip

Another useful fact is \emph{Robin's inequality} for the $\sigma$ function
(sum of the divisors):
\begin{equation}
  \sigma(n) < e^{\gamma} n \log \log n \, , \hspace{2em} \text{$n$ large,}
\end{equation}
where $\gamma$ is Euler's constant. This is true for large $n$,
where large means $n \ge 5041$, and if and only if Riemann's hypothesis is
true~\cite{Robin:1984}.

\begin{example}
  The rightmost pole of the Dirichlet series $F^\triangle(s)$ in
  Equation~\eqref{eq:Dirichlet-triangle} is found for $s=2$, has order
  $1$ and its residue is $\zeta(2)/6$. Using the above theorem we
  conclude that the partial sum of the terms in the sequence
  $f^\triangle$ behaves asymptotically as
  \begin{equation}
    \boxed{\sum_{n=1}^N f^\triangle(n) \sim \frac{\zeta(2)}{12} N^2 = \frac{\pi^2}{72} N^2} \, ,  
  \end{equation}
  and the sequence itself grows asymptotically as $f^{\triangle}(N) =
  \mathcal{O}(N)$.
  
  For large $n$, the leading term is $\zeta(s) \zeta(s-1) / 6$, hence
  \begin{equation}
    \boxed{f^\triangle(n) < \frac{e^\gamma n \log \log n}{6} \, , \hspace{2em} \text{$n$ large.}}    
  \end{equation}
\end{example}

\section{Examples}
\label{sec:examples}

\subsection{Orbifolds of the conifold}
\label{sec:square}

Another simple geometry that lends itself to the counting of its
orbifolds is the \emph{conifold}. In terms of the dimer model
description, it corresponds to the bipartite square lattice with a
black and a white vertex in its unit cell. The brane tilings for the
first orbifolds can be found in Table~\ref{tab:Conifold-tilings}. Note
that as mentioned earlier, we are here enumerating all the toric
geometries stemming from orbifolds of the conifold and \emph{not} all
possible quiver gauge theories, since we are not taking into account
the multiple toric phases.

\begin{table}
  \centering
  \begin{tabular}{SSSSS}
    \toprule
    $n$ & $ 1$ & $2$ & $2$ & $3$ \\ \midrule
    brane tiling & \includegraphics[width=.17\textwidth]{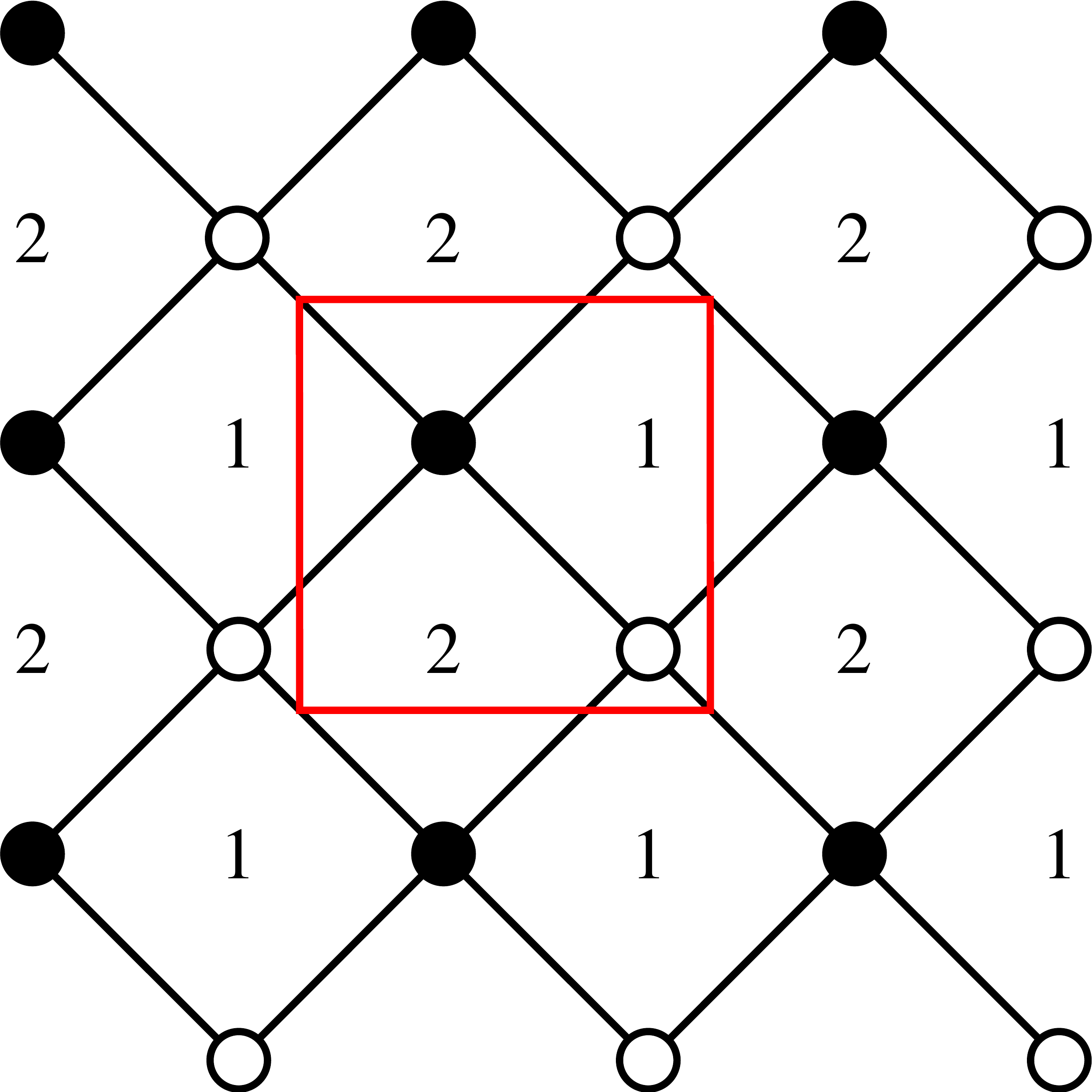} & \includegraphics[width=.17\textwidth]{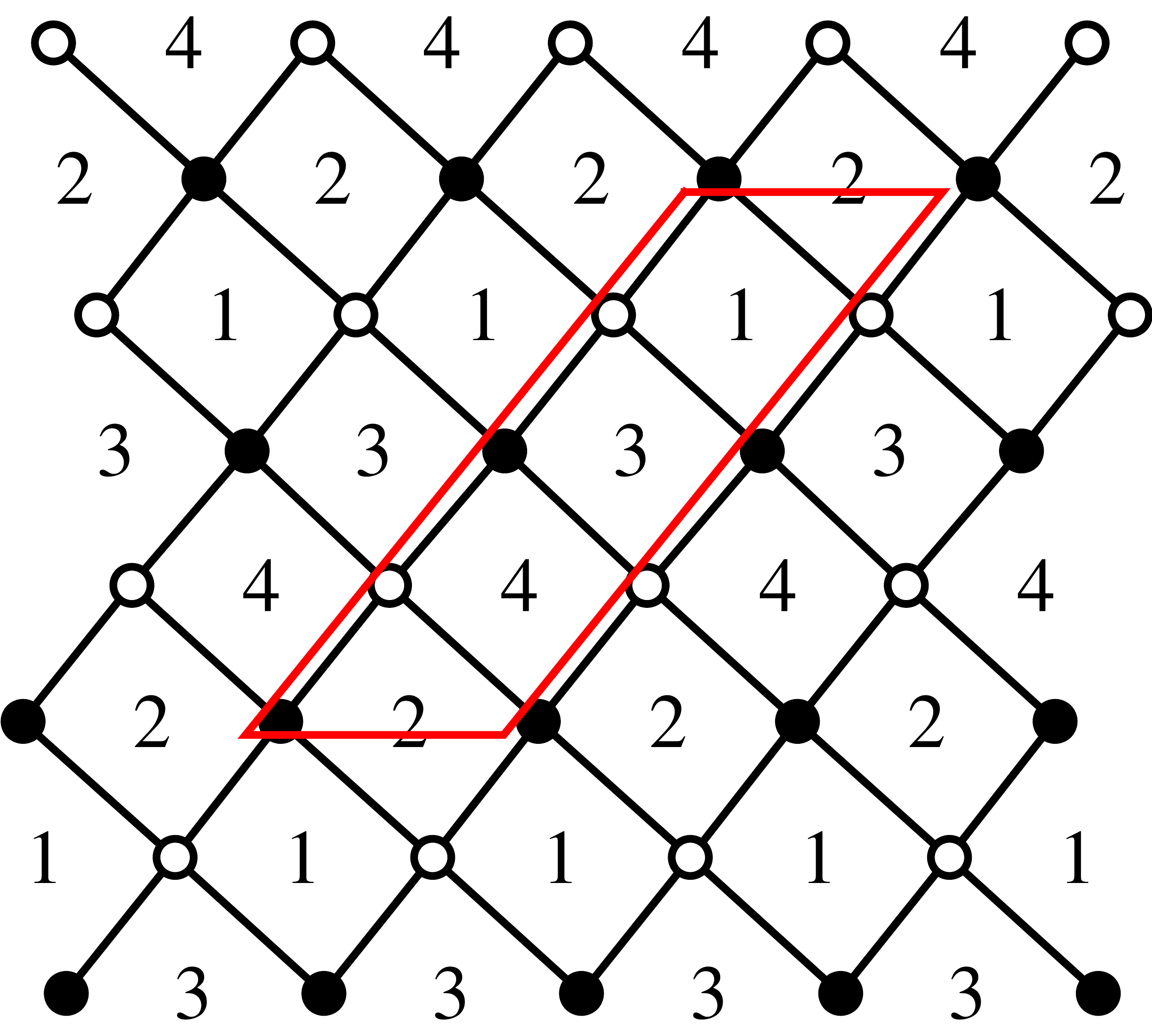} & \includegraphics[width=.17\textwidth]{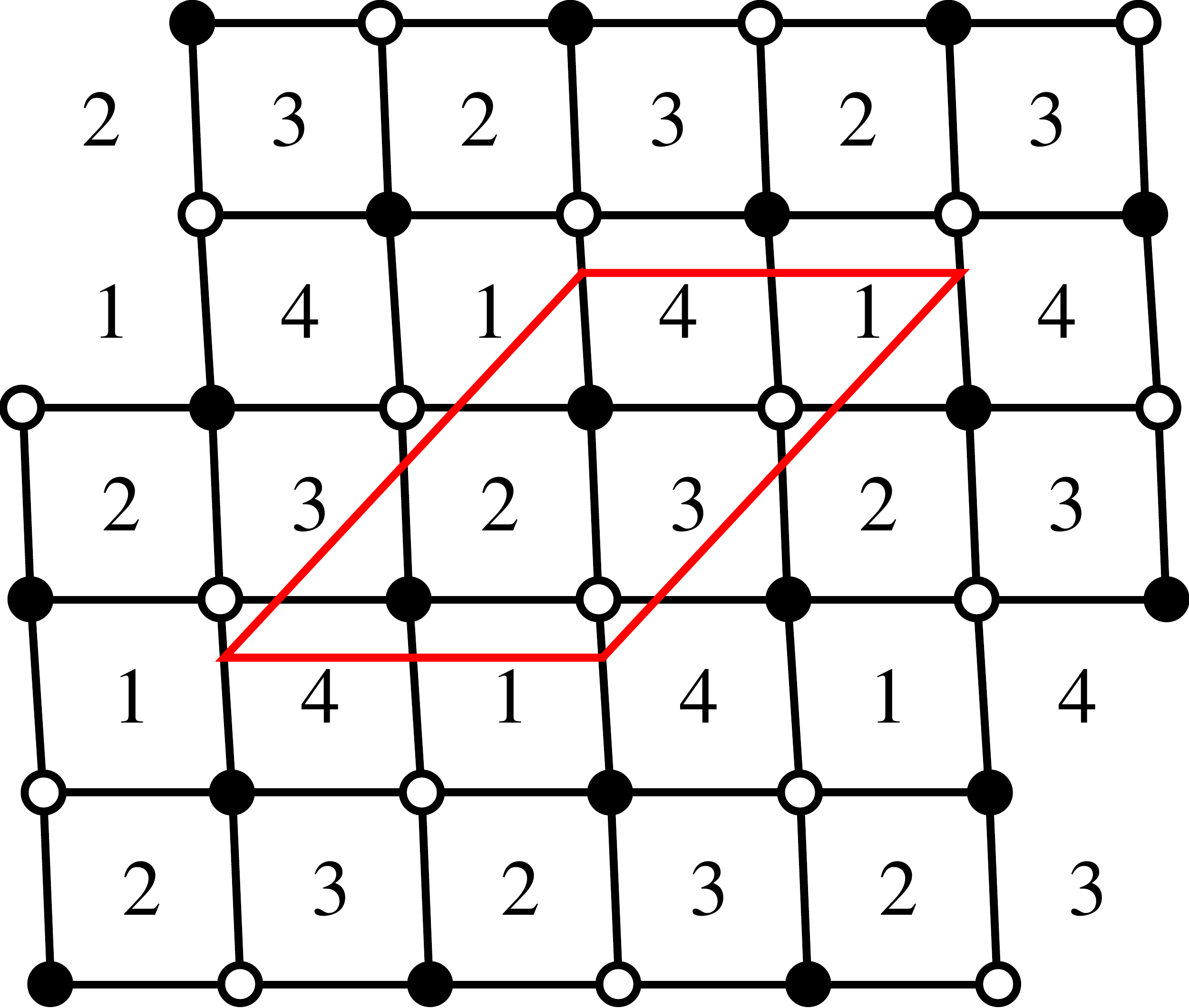}  & \includegraphics[width=.17\textwidth]{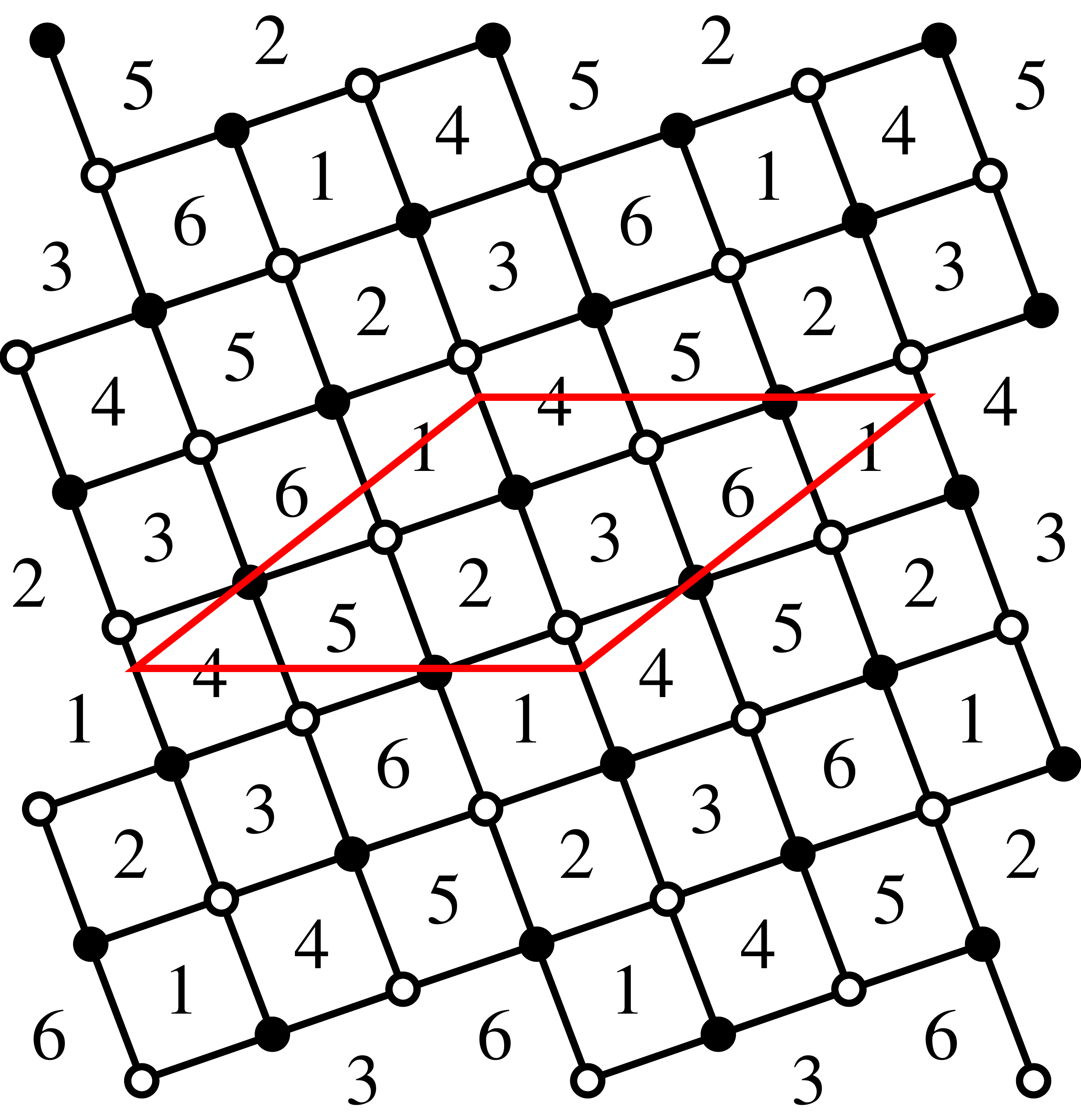}  \\
    geometry & conifold & $L^{222}$ & $\mathbb{F}_0$ & $Y^{3,0}$ \\ \bottomrule
  \end{tabular}
  \caption{Brane tilings for the first orbifolds of the conifold.}
  \label{tab:Conifold-tilings}
\end{table}

\begin{figure}
  \centering
  \includegraphics[width=.8\textwidth]{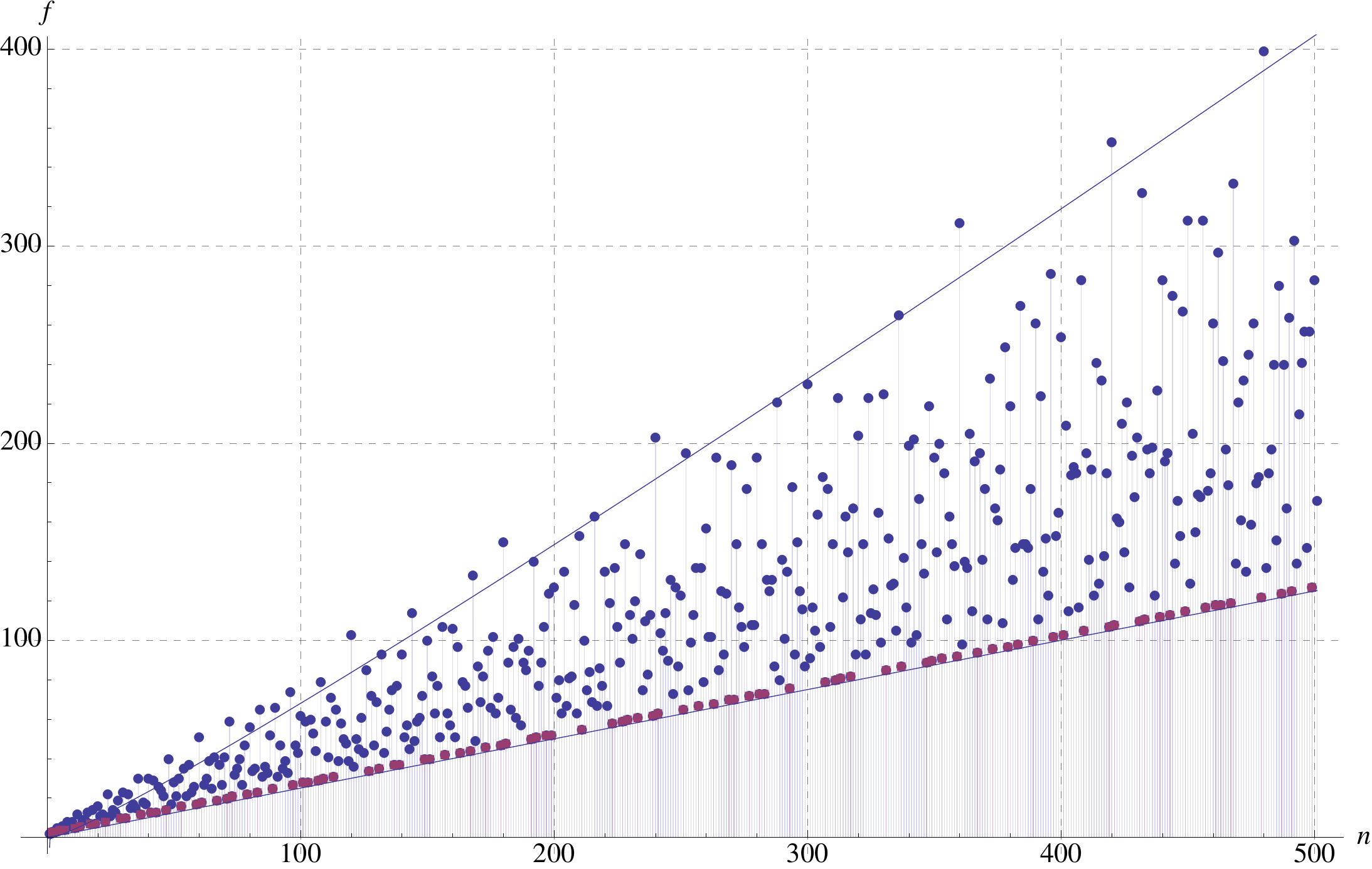}
  \caption{Scatter plot of the sequence $f^\square$ for a square
    lattice. Prime numbers are given in red. The two lines correspond to
    $n/4 $ and $e^\gamma n \log \log n / 4 $.}
  \label{fig:square-scatter}
\end{figure}

The brane tiling corresponding to the conifold is a bipartite square
lattice with two faces in the unit cell. Its symmetry is given by Klein's Vierergruppe $V = \setZ_2
\times \setZ_2$, with cycle index
\begin{equation}
  Z_V = \frac{1}{4} \left( x_1^4 + 2\, x_1^2 x_2 + x_2^2 \right) \, .  
\end{equation}
Accordingly, we can decompose $f^\square$ as in
Table~\ref{tab:square-numbers-V4}.  The numbers $f^\square (n) $, $ n
\le 500$ are represented in Figure~\ref{fig:square-scatter}.
\begin{table}
  \centering
  \begin{tabular}{ccccccccccccccccccccc}
    \toprule 
     & 1 & 2 & 3 & 4 & 5 & 6 & 7 & 8 & 9 & 10 & 11 & 12 & 13 & 14 & 15 & 16 \\
    \midrule
    $f^{\square}_{x_1^4}$ & 1 & 3 & 4 & 7 & 6 & 12 & 8 & 15 & 13 & 18 & 12 & 28 & 14 & 24 & 24 & 31 \\
    $f^{\square}_{x_1^2 x_2}$ & 1 & 1 & 1 & 2 & 2 & 1 & 1 & 3 & 2 & 2 & 1 & 3 & 2 & 1 & 2 & 4 \\
    $f^{\square}_{x_2^2}$ & 1 & 3 & 2 & 5 & 2 & 6 & 2 & 7 & 3 & 6 & 2 & 10 & 2 & 6 & 4 & 9 \\
    \midrule
    $f^{\square}$ & 1 & 2 & 2 & 4 & 3 & 5 & 3 & 7 & 5 & 7 & 4 & 11 & 5 & 8 & 8 & 12 \\
    \bottomrule
  \end{tabular}
  \caption{Number of sublattices of index $n$ for the square lattice, classified by the cycles of the Vierergruppe $V = \setZ_2 \times \setZ_2$.}
  \label{tab:square-numbers-V4}
\end{table}

The analysis of the sequences corresponding to the terms in $Z_V$
gives the following results.
\begin{enumerate}
\item For the monomial $x_1^4$ we obtain the by now usual term
  \begin{equation}
    f^\square_{x_1^4} = \un * \N \, .    
  \end{equation}
  This is the same as for the example of the hexagon since it depends
  only on the dimension of the lattice.
\item The monomial $x_1^2 x_2$ receives two contributions (this is
  different from the other cases we have encountered):
  \begin{equation}
    f^\square_{x_1^2 x_2} = \frac{1}{2} \un * \left( \chi_{4,2} +  \set{1,-1,0,2} * \un \right) \, .    
  \end{equation}
  The corresponding power series is given by
  \begin{equation}
    G^\square_{x_1^2 x_2} = \frac{1}{\left(1 - t \right) \left( 1 + t^2 \right)} - 1 \, .    
  \end{equation}
\item The monomial $x_2^2$ corresponds to
  \begin{equation}
    f^\square_{x_2^2} = \set{1,1} * \un * \un \, .    
  \end{equation}
  and the corresponding power series is
   \begin{equation}
     G^\square_{x_2}(t) = \frac{1-t^3}{\left( 1 - t \right)\left( 1-t^2 \right)} - 1 \, .
  \end{equation}
\end{enumerate}
The generating functions in this decomposition are collected in
Table~\ref{tab:Dirichlet-square-V4}.

\begin{table}
  \centering
  \begin{tabular}{cccc}
    \toprule
    symmetry  & Dirichlet series $G(s)$ & generating function $G(t)$ \\
    \midrule
    $x_1^4$ & $\zeta(s-1) $ & $\frac{1+t^3}{(1-t)(1-t^2)} - 1$\\
    $x_1^2 x_2 $ & $\frac{1}{2}\left( L(s, \chi_{4,2} )+ \left( 1 - 2^{-s} + 2^{1-2s} \right) \zeta(s) \right)$  & $\frac{1}{(1-t)(1+t^2)} - 1$\\
    $x_2^2 $ & $\left( 1 + 2^{-s} \right) \zeta(s)$ & $\frac{1-t^3}{(1-t)(1-t^2)} - 1$ \\
    \bottomrule
  \end{tabular}
  \caption{Generating functions for the Abelian orbifolds of the conifold organized by the symmetries of the brane tiling.}
  \label{tab:Dirichlet-square-V4}
\end{table}

\bigskip 

Collecting all the terms and summing them according to the coefficients
of the cycle index, we find 
\begin{enumerate}
\item For the power series:
  \begin{equation}
    G^\square(t) =  \frac{1}{\left( 1 - t \right) \left( 1 - t^4 \right)} - 1 \, ,
  \end{equation}
  whence
  \begin{equation}
    \label{eq:power-square}
    \boxed{F^\square(t) = \sum_{m=1}^\infty \left[ \frac{1}{\left( 1 - t^m \right) \left( 1 - t^{4m} \right)} - 1 \right] = \sum_{m=1}^\infty \sum_{\substack{n_1, n_2= 0 \\ \neq \left( 0, 0 \right)}}^\infty  t^{m \left( n_1 + 4 n_2  \right)}} \, .
  \end{equation}
\item For the Dirichlet series:
  \begin{equation}
    \label{eq:Dirichlet-square}
    \boxed{F^\square(s) = \frac{1}{4}\, \zeta(s) \left[ \zeta(s-1) + 2 \left( 1 + 2^{-2s} \right) \zeta(s) +  L(s,\chi_{4,2}) \right] } \, .
  \end{equation}
\end{enumerate}

The asymptotic behavior of the partial sum of the terms in the
sequence $f^\square$ is given by \begin{equation} \sum_{n=1}^N
  f^\square(n) \sim \frac{\zeta(2)}{8} N^2 = \frac{\pi^2}{48} N^2 \, ,
\end{equation}
and the sequence itself grows asymptotically as $f^{\square}(N) =
\mathcal{O}(N)$. By Robin's inequality, the sequence $f^\square$ is bounded for large $n$ by
\begin{equation}
  f^\square(n) < \frac{e^\gamma n \log \log n}{4} \, , \hspace{2em} \text{$n$ large.}
\end{equation}

\subsection{Orbifolds of the $L^{aba}$ theories}
\label{sec:orbif-laba-theor}

\begin{table}
  \centering
  \begin{tabular}{ccccccccccccccccccccc}
    \toprule 
     & 1 & 2 & 3 & 4 & 5 & 6 & 7 & 8 & 9 & 10 & 11 & 12 & 13 & 14 & 15 & 16 \\
    \midrule
    $f^{\trap}_{x_1^2}$ & 1 & 3 & 4 & 7 & 6 & 12 & 8 & 15 & 13 & 18 & 12 & 28 & 14 & 24 & 24 & 31 \\
    $f^{\trap}_{x_2}$   & 1 & 1 & 2 & 3 & 2 & 2 & 2 & 5 & 3 & 2 & 2 & 6 & 2 & 2 & 4 & 7 \\ \midrule
    $f^{\trap}$        & 1 & 2 & 3 & 5 & 4 & 7 & 5 & 10 & 8 & 10 & 7 & 17 & 8 & 13 & 14 & 19 \\
    \bottomrule
  \end{tabular}
  \caption{Number of sublattices of index $n$ for the $L^{aba}$ theories ($\setZ_2 $ symmetry).}
  \label{tab:Laba-numbers}
\end{table}

In the case of the $L^{aba}$ theories with $a \neq b$, the brane
tiling lattice has $\setZ_2 $ symmetry, and the cycle index is given by
\begin{equation}
  Z_{\setZ_2} = \frac{1}{2} \left( x_1^2 + x_2 \right) \, .  
\end{equation}
The number of orbifolds $f^\trap(n)$ (the first terms are collected in
Table~\ref{tab:Laba-numbers}) can be decomposed into two
contributions:
\begin{enumerate}
\item the usual term corresponding to the identity,
  \begin{equation}
    f^\trap_{x_1^2} = \un * \N \, ;
  \end{equation}
\item the term corresponding to the reflection $x_2$:
  \begin{equation}
    f^\trap_{x_2} = \set{1,-1,0,2} * \un * \un \, .    
  \end{equation}
\end{enumerate}
The power series reads:
\begin{equation}
  \boxed{ F^\trap(t) = \sum_{m=1}^\infty \left[ \frac{1+t^{3m}}{\left( 1 - t^m \right) \left( 1 - t^{4m} \right) } \right]  = \sum_{m=1}^\infty \sum_{\substack{n_1, n_2= 0 \\ \neq \left( 0, 0 \right)}}^\infty  t^{m \left( n_1 + 4 n_2  \right)} \left( 1 + t^{3m} \right) } \, .
\end{equation}
The Dirichlet series is 
\begin{equation}
  \boxed{F^\trap(s) = \frac{1}{2} \zeta(s) \left[ \zeta(s-1) + \left( 1 - 2^{-s} + 2^{1-2s}  \right) \zeta(s) \right] }\, .  
\end{equation}
From here we can read the asymptotic behavior of the partial sum of
the terms in the sequence $f^\trap$:
\begin{equation}
  \sum_{n=1}^N f^\trap(n) \sim \frac{\zeta(2)}{4} N^2 =
  \frac{\pi^2}{24} N^2 \, .
\end{equation}
By Robin's inequality, the sequence $f^\trap$ is bounded for large
$n$ by
\begin{equation}
  f^\trap(n) < \frac{e^\gamma n \log \log n}{2} \, , \hspace{2em} \text{$n$ large.}
\end{equation}

If $a=b$, the brane tiling acquires an extra symmetry and the relevant
group is $\setZ_2 \times \setZ_2$. Since this is the same symmetry as
for the brane tiling of the conifold that has been described in the
previous section, the formulae in Equation~\eqref{eq:power-square} and
Equation~\eqref{eq:Dirichlet-square} apply also to this case.

\subsection{Orbifolds of $\setC^4$}
\label{sec:tetrahedron}

\begin{figure}
  \centering
  \includegraphics[width=5cm]{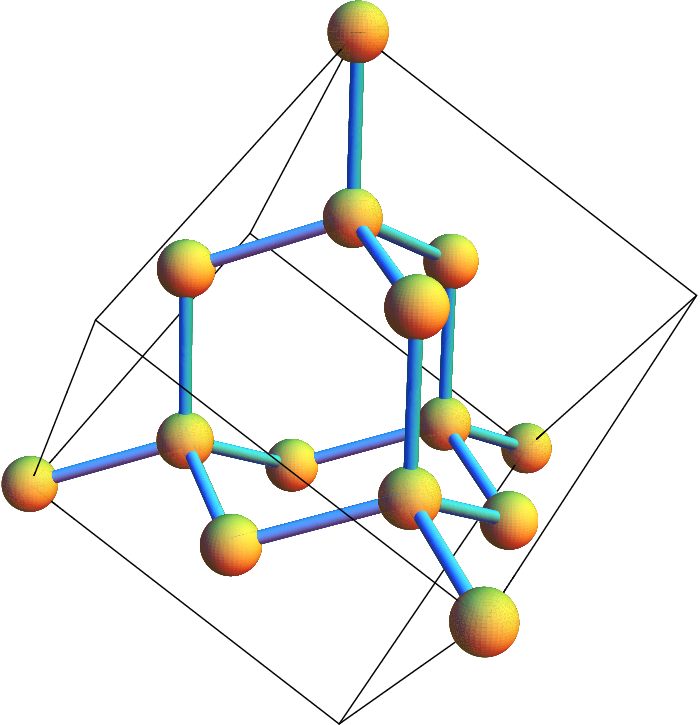}
  \caption{The tetrahedral lattice. Points are at the vertices and
    the center of a tetrahedron. Four such tetrahedra connected at their
    vertices fit into a cube.}
  \label{fig:tetrahedron}
\end{figure}

For the last example we consider Abelian orbifolds of $\setC^4$. The
three--dimensional counterpart of the brane tiling was described
in~\cite{Lee:2007kv}. The bipartite lattice has $S^4$ symmetry, like a
tetrahedral lattice (see Figure~\ref{fig:tetrahedron}). For this
reason we will denote the counting function $f^\tetra (n)$ by a
tetrahedron, just as before we used a triangle for the $S_3$ symmetry
of the orbifolds of $\setC^3$.

\begin{figure}
  \centering
  \includegraphics[width=.8\textwidth]{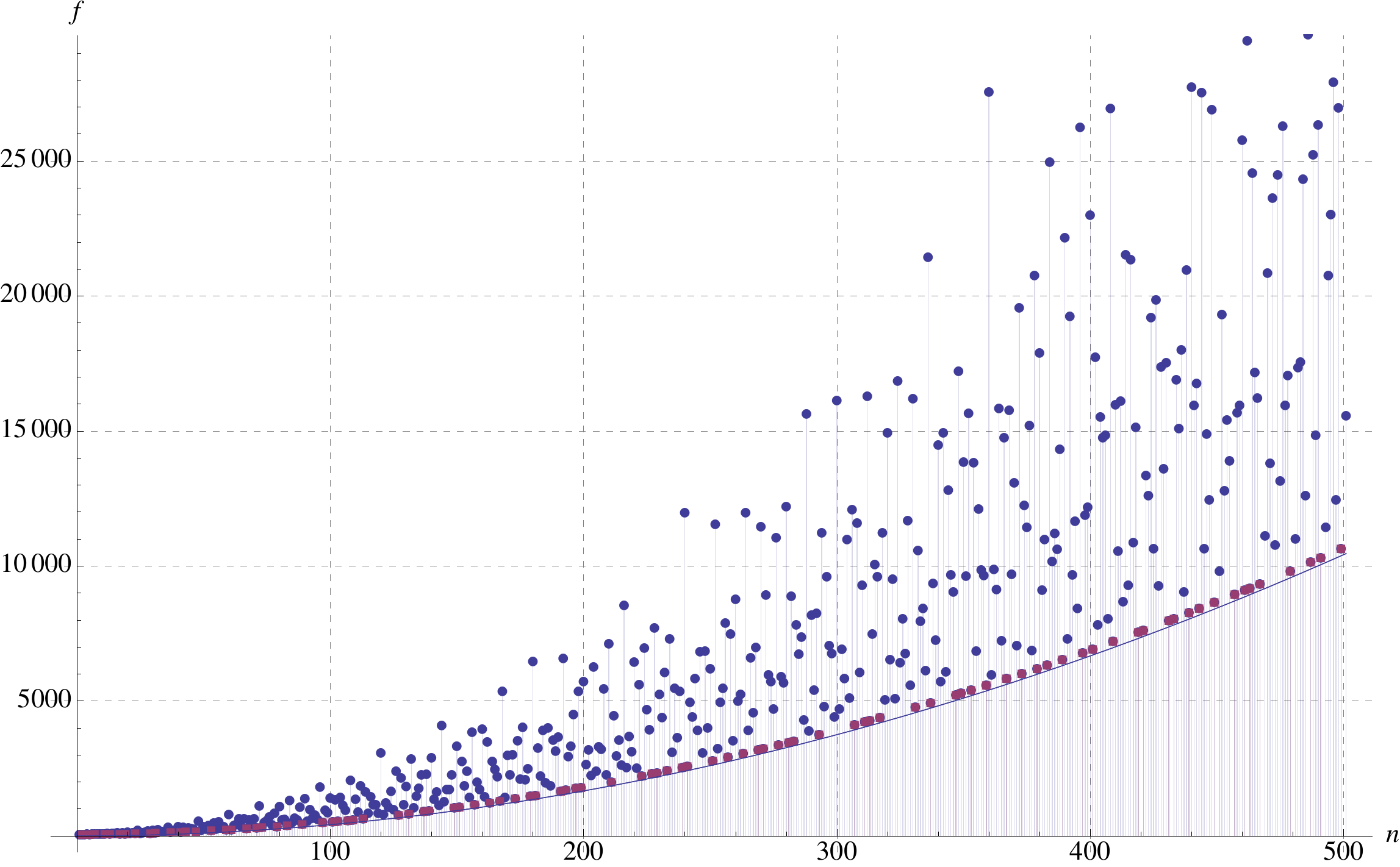}
  \caption{Scatter plot of the number of sublattices of index $n$ for
    a tetrahedral lattice. Prime numbers are given in red. The line corresponds to
    $n^2/24 $.}
  \label{fig:tetrahedron-scatter}
\end{figure}

The cycle index for $S_4$ is
\begin{equation}
  Z_{S_4} = \frac{1}{24} \left( x_1^4 + 6\, x_1^2 x_2 + 3\, x_2^2 + 8\, x_1 x_3 + 6\, x^4 \right) \, . 
\end{equation}
The first terms of the subsequences
$f_{\mathbf{x}^\alpha}^{\tetra}(n)$ are collected in
Table~\ref{tab:tetrahedron-numbers}, and the first 500 numbers of
$f^{\tetra }(n)$ are represented in
Figure~\ref{fig:tetrahedron-scatter}.

\begin{table}
  \centering
  \begin{tabular}{ccccccccccccccccccccc}
    \toprule 
     & 1 & 2 & 3 & 4 & 5 & 6 & 7 & 8 & 9 & 10 & 11 & 12 & 13 & 14 & 15 & 16 \\
    \midrule
    $f^{\tetra}_{x_1^4}$ & 1 & 7 & 13 & 35 & 31 & 91 & 57 & 155 & 130 & 217 & 133 & 455 & 183 & 399 & 403 & 651  \\
    $f^{\tetra}_{x_1^2 x_2}$ & 1 & 3 & 5 & 11 & 7 & 15 & 9 & 31 & 18 & 21 & 13 & 55 & 15 & 27 & 35 & 75 \\
    $f^{\tetra}_{x_2^2}$ & 1 & 3 & 5 & 11 & 7 & 15 & 9 & 31 & 18 & 21 & 13 & 55 & 15 & 27 & 35 & 75 \\
    $f^{\tetra}_{x_1 x_3}$ & 1 & 1 & 1 & 2 & 1 & 1 & 3 & 2 & 4 & 1 & 1 & 2 & 3 & 3 & 1 & 3 \\
    $f^{\tetra}_{x_4}$ & 1 & 1 & 1 & 3 & 3 & 1 & 1 & 5 & 2 & 3 & 1 & 3 & 3 & 1 & 3 & 7 \\
    \midrule
    $f^{\tetra}$ & 1 & 2 & 3 & 7 & 5 & 10 & 7 & 20 & 14 & 18 & 11 & 41 & 15 & 28 & 31 & 58 \\
    \bottomrule
  \end{tabular}
  \caption{Number of sublattices of index $n$ for the tetrahedral lattice, classified by the cycles of the symmetric group $S_4$.}
  \label{tab:tetrahedron-numbers}
\end{table}

\bigskip

Following equation \eqref{eq:fL3-as-convolution}, we know that each of
the subsequences can be written in the form of a double Dirichlet
convolution where one of the factors is the unity $\un$:
\begin{enumerate}
\item The sequence $f^{\tetra}_{x_1^4}$ is the one corresponding to
  the identity. This means that it is given by
  Equation~(\ref{eq:generic-lattice-f(n)}) for $d=3$. Equivalently, it can
  be written as the convolution
  \begin{equation}
    f^{\tetra}_{x_1^4} = \un * \N * \N^2 \, .    
  \end{equation}
\item The sequence $f^{\tetra}_{x_1^2 x_2} $ can be written as
  \begin{equation}
    f^{\tetra}_{x_1^2 x_2} = \set{1, -1, 0, 4} * \un * \un * \N \, .
  \end{equation}
\item The sequence $f^{\tetra}_{x_2^2} $ coincides with
  $f^{\tetra}_{x_1^2 x_2}$.
\item The sequence $f^{\tetra}_{x_1 x_3}$ can be written as a
  convolution with the non--principal Dirichlet character of modulus
  three that has already appeared before:
  \begin{equation}
    f^{\tetra}_{x_1 x_3} = \set{1,0,-1,0,0,0,0,0,3} * \un * \un * \chi_{3,2} \, .    
  \end{equation}
\item The sequence $f^{\tetra}_{x_4} $ is the convolution of the
  non--principal character of modulus four:
  \begin{equation}
    f^{\tetra}_{x_4} = \set{1,-1,0,2} * \un * \un * \chi_{4,2} \, .    
  \end{equation}
\end{enumerate}
We already have the Dirichlet series corresponding to each of these
terms and we can collect them by using Burnside's lemma. The final
result is:
\begin{multline}
  F^{\tetra}(s) = \frac{1}{24} \left( F^{\tetra}_{x_1^4}(s) _ + 6\, F^{\tetra}_{x_1^2 x_2}(s) + 3\, F^{\tetra}_{x_2^2} + 8\, F^{\tetra}_{x_1 x_3}(s) + 6\, F^{\tetra}_{x^4}(s) \right) = \\
  = \frac{\zeta(s) \zeta(s-1)}{24} \left( \zeta(s-2) + 9 \left( 1
      - 2^{-s} + 2^{2-2s} \right) \zeta(s)\right) + \\ +
  \frac{\zeta(s)^2}{24} \left( 8 \left( 1 - 3^{-s} + 3^{1-2s}
    \right) L(\chi_{3,2},s) + 6 \left( 1 - 2^{-s} + 2^{1-2s}
    \right) L(\chi_{4,2},s) \right) \, .
\end{multline}
The rightmost pole is at $s=3$ and has order $1$. This means that the
partial sum of the terms in the sequence behaves like
\begin{equation}
  \sum_{n=1}^N f^{\tetra}(n) \sim \frac{ \zeta(2) \zeta(3)}{3 \times 24} N^3 \sim 0.0274 N^3 \, .
\end{equation}
It is also possible to write the power series corresponding to each
term as follows:
\begin{equation}
  F^{\tetra}_{\mathbf{x}^\alpha }(t) = \sum_{m=1}^\infty G^{\tetra}_{\mathbf{x}^\alpha } (t^m) \, ,  
\end{equation}
where
\begin{subequations}
  \begin{gather}
    G^{\tetra}_{x_1^4} (t) = \sum_{n,m=1}^\infty n m^2 t^{m n} \, , \\
    G^{\tetra}_{x_1^2 x_2} (t) = G^{\tetra}_{x_2^2} (t) = \sum_{n,m=1}^\infty m \left( t^{m n } - t^{ 2 m n} + 4 t^{ 4 m n} \right) \, ,\\
    G^{\tetra}_{x_1 x_3} (t) = \frac{1}{2}  \left[ \sum_{n,m=-\infty}^\infty t^{ n^2 + 4 m^2} - 1 \right] \, ,\\
    G^{\tetra}_{x_4} (t) = \frac{1}{2}  \left[ \sum_{n,m=-\infty}^\infty t^{n^2 + m n + 7 m^2 } - 1 \right] \, .\\
  \end{gather}
\end{subequations}

\begin{table}
  \centering
  \begin{tabular}{cccc}
    \toprule
    symmetry & Dirichlet series $G(s)$ & generating function $F(t)$ \\
    \midrule
    $x_1^4$  & $ \zeta(s-1) \zeta(s-2)$ & $ \displaystyle{ \sum_{n,m=1}^\infty n m^2 t^{m n}} $\\
    $x_1^2 x_2$ & $\left( 1 - 2^{-s} + 2^{2-2s} \right) \zeta(s) \zeta(s-1)$ & $\displaystyle{ \sum_{n,m=1}^\infty m \left( t^{m n } - t^{ 2 m n} + 4 t^{ 4 m n} \right)}$\\
    $x_2^2$  & $\left( 1 - 2^{-s} + 2^{2-2s} \right) \zeta(s) \zeta(s-1)$ & $\displaystyle{\sum_{n,m=1}^\infty m \left( t^{m n } - t^{ 2 m n} + 4 t^{ 4 m n} \right)}$\\
    $x_1 x_3$ & $\left( 1 - 3^{-s} + 3^{1-2s} \right) L(s,\chi_{3,2}) \zeta(s)$ & $\frac{1}{2} \displaystyle{ \left[ \sum_{n,m=-\infty}^\infty t^{ n^2 + 4 m^2} - 1 \right]} $\\
    $x_4$ & $\left( 1 - 2^{-s} + 2^{1-2s} \right) L(s,\chi_{4,2}) \zeta(s)$ & $ \frac{1}{2} \displaystyle{ \left[ \sum_{n,m=-\infty}^\infty t^{n^2 + m n + 7 m^2 } - 1 \right]}$ \\
    \bottomrule
  \end{tabular}
  \caption{Generating functions for the Abelian orbifolds of $\setC^4$, organized by the symmetries of the brane tiling.}
  \label{tab:Dirichlet-tetrahedron}
\end{table}

\subsection*{Acknowledgements}

We would like to thank the organizers of the workshop ``Branes,
Strings and Black Holes'' at the Yukawa Institute for Theoretical
Physics (Kyoto) where this work was initiated.

A.~H.~ would like to thank Rak-Kyeong Seong and John Davey for collaboration on results which were used in this paper, and to Yang-Hui He, Giuseppe Torri, and Noppadol Mekareeya for useful discussions.

The research of D.~O. and S.~R. was supported by the World Premier
International Research Center Initiative (WPI Initiative), MEXT,
Japan.

\bibliography{SubReferences}

\providecommand{\href}[2]{#2}\begingroup\raggedright\begin{thebibliography}{10}

\bibitem{Hanany:2005ve}
A.~Hanany and K.~D. Kennaway, {\it {Dimer models and toric diagrams}},
  \href{http://xxx.lanl.gov/abs/hep-th/0503149}{ hep-th/0503149}.

\bibitem{Franco:2005rj}
S.~Franco, A.~Hanany, K.~D. Kennaway, D.~Vegh, and B.~Wecht, {\it {Brane Dimers
  and Quiver Gauge Theories}},  {\em JHEP} {\bf 01} (2006) 096,
  [\href{http://xxx.lanl.gov/abs/hep-th/0504110}{ hep-th/0504110}].

\bibitem{Hanany:2008cd}
A.~Hanany and A.~Zaffaroni, {\it {Tilings, Chern-Simons Theories and M2
  Branes}},  {\em JHEP} {\bf 10} (2008) 111,
  [\href{http://xxx.lanl.gov/abs/0808.1244}{ 0808.1244}].

\bibitem{Davey:2009bp}
J.~Davey, A.~Hanany, and J.~Pasukonis, {\it {On the Classification of Brane
  Tilings}},  \href{http://xxx.lanl.gov/abs/0909.2868}{ 0909.2868}.

\bibitem{Kucab:1981}
M.~Kucab, {\it Colour lattices and spin translation groups. general case},
  {\em Acta Cryst. A} {\bf 37} (1981) 17--21.

\bibitem{Rutherford:1904p2035}
J.~S. Rutherford, {\it The enumeration and symmetry-significant properties of
  derivative lattices},  {\em Acta Cryst. A} {\bf 48} (1992) 500--508.

\bibitem{Baake:1997a}
M.~Baake, {\it Solution of the coincidence problem in dimensions $d\le 4$},  in
  {\em The Mathematics of Long-Range Aperiodic Order} (R.~Moody, ed.),
  pp.~9--44.
\newblock Kluwer, Dordrecht, 1997.
\newblock \href{http://xxx.lanl.gov/abs/math/0605222}{ math/0605222}.

\bibitem{Rutherford:2009p1973}
J.~S. Rutherford, {\it Sublattice enumeration. iv. equivalence classes of plane
  sublattices by parent patterson symmetry and colour lattice group type},
  {\em Acta Cryst. A} {\bf 65} (2009) 156--163.

\bibitem{Rak}
J.~Davey, A.~Hanany, and R.-K. Seong, {\it {Counting Orbifolds}},
  \href{http://xxx.lanl.gov/abs/1002.****}{ 1002.****}.

\bibitem{Biggs:2003}
N.~L. Biggs, {\em Discrete Mathematics}, ch.~27.
\newblock Oxford University Press, 2003.

\bibitem{Apostol:1998}
T.~M. Apostol, {\em Introduction to analytic number theory}.
\newblock Undergraduate Texts in Mathematics. Springer-Verlag, New
  York--Heidelberg, 1976.

\bibitem{Robin:1984}
G.~Robin, {\it Grandes valeurs de la fonction somme des diviseurs et hypothèse
  de riemann},  {\em J. Math. Pures Appl. (9)} {\bf 63} (1984), no.~2 187--213.

\bibitem{Lee:2007kv}
S.~Lee, S.~Lee, and J.~Park, {\it {Toric AdS(4)/CFT(3) duals and M-theory
  crystals}},  {\em JHEP} {\bf 05} (2007) 004,
  [\href{http://xxx.lanl.gov/abs/hep-th/0702120}{ hep-th/0702120}].

\end{thebibliography}\endgroup

\newpage

\appendix

\section{Generic lattice in $d$ dimensions}
\label{sec:generic-lattice-d}

In this appendix we derive a general expression for the number
$f^{L_d}(n)$ of sublattices of index $n$ for a generic lattice in $d$
dimensions. Using the construction in
Section~\ref{sec:hermite-normal-form}, we want to count matrices of the
form
\begin{equation}
  \begin{pmatrix}
    a_{11} \\
    a_{21} & a_{22} \\
    \dots & & \ddots \\
    a_{d1} & a_{d2} & \dots & a_{dd}
  \end{pmatrix}
\end{equation}
where $a_{11} a_{22} \cdots a_{dd} = n$, and $0 \le a_{ij} < a_{ii}
\, , \forall j < i$.

\begin{itemize}
\item For $d=1$, the only condition is $a_{11} = n$, so there is
  exactly one sublattice for each choice of the index:
  \begin{equation}
    f^{L_1} (n) = 1 \, .
  \end{equation}
\item For $d=2$ we want to count the triangular matrices
  \begin{equation}
    \begin{pmatrix}
      a_{11} & 0 \\
      a_{21} & a_{22} 
    \end{pmatrix} 
  \end{equation}
  such that $a_{11} a_{22} = n $ and $a_{21} = 0, 1, \dots ,
  a_{22} - 1$. It is immediate to see that there are $m$ such matrices for
  each choice of a divisor $m$ of $n$:
  \begin{equation}
    f^{L_2} (n) = \sum_{m|n} m = \sum_{m|n} m f^{L_1} (m) \, .  
  \end{equation}
\item In $d=3$ we count the matrices
  \begin{equation}
    \begin{pmatrix}
      a_{11} & 0 & 0 \\
      a_{21} & a_{22} & 0 \\
      a_{31} & a_{32} & a_{33}
    \end{pmatrix}
  \end{equation}
  where $a_{11} a_{22} a_{33} = n $, $a_{21} = 0 , 1 , \dots , a_{22}
  - 1 $ and both $a_{31} $ and $a_{32} $ are in $\set{0, 1, \dots,
    a_{33} - 1}$. Let us write
  \begin{align}
    a_{11} = \frac{n}{m_1} \, , && a_{22} = \frac{m_1}{m_2} \, ,&& a_{33} = m_2 \, ,  
  \end{align}
  then we have $a_{22} a_{33}^2 = m_1 / m_2 \times m_2^2 = m_1 m_2 $
  such matrices for any divisor $m_1$ of $n$ and any divisor $m_2$ of
  $m_1$. Hence
  \begin{equation}
    f^{L_3} (n) = \sum_{m_1 | n} \sum_{m_2| m_1} m_1 m_2 = \sum_{m|n} m f^{L_2}(m) \, .
  \end{equation}
\item For general $d$, one has the condition $a_{11} a_{22} \dots
  a_{dd} = n$ and a total of $a_{22} a_{33}^2 \cdots a_{dd}^{d-1}$
  matrices. Writing the coefficients as
  \begin{align}
    a_{11} = \frac{n}{m_1} \, , && a_{22} = \frac{m_1}{m_2} \, , && \dots &&
    a_{(d-1)(d-1)} = \frac{m_{d-2}}{m_{d-1}} \, , && a_{dd} = m_{d-1} \, ,
  \end{align}
  we obtain
  \begin{equation}
    f^{L_d} (n) = \sum_{m_1 | n } \sum_{m_2| m_1} \dots \sum_{m_{d-1}| m_{d-2}} m_1 m_2 \cdots m_{d-1} = \sum_{m|n} m f^{L_{d-1}} (m) \, .
  \end{equation}
\end{itemize}

The generating functions for these sequences are written as
follows. From the observation that there are $a_{22} a_{33}^2 \cdots
a_{dd}^{d-1} $ matrices of index $n = a_{11} a_{22} \cdots a_{dd}$ we
can obtain directly the power series
\begin{equation}
  \label{eq:generic-lattice-power-series}
  F^{L_d} (t) = \sum_{n=1}^\infty f^{L_d} (n) t^n = \sum_{k_1, k_2, \dots k_d =1}^\infty k_2 k_3^2 \cdots k_d^{d-1} t^{k_1 k_2 \cdots k_d} \, .
\end{equation}
The Dirichlet series can be written by using the recursion relation
\begin{equation}
  f^{L_d} (n) = \sum_{m|n} m f^{L_{d-1}} (m) \, .
\end{equation}
Rewriting it as
\begin{equation}
  \frac{f^{L_d} (n)}{n} = \sum_{m|n} \frac{1}{n/m} f^{L_{d-1}} (m) \, ,
\end{equation}
we see that the sequence $[ f^{L_d}(n) / n] $ is the Dirichlet
convolution of the sequences $[1/n]$ and $f^{L_{d-1}}$:
\begin{equation}
  [ \frac{f^{L_d}(n)}{n} ] = [ \frac{1}{n}] * [f^{L_{d-1}} (n) ] \, .
\end{equation}
It follows that
\begin{equation}
  F^{L_d} ( s + 1 ) = \zeta(s+1) F^{L_{d-1}} (s) \, .
\end{equation}
This recursion relation can be solved starting from
\begin{equation}
  F^{L_1} (s) = \zeta (s) \, ,
\end{equation}
and we find
\begin{equation}
  F^{L_d}(s) = \zeta(s) \zeta(s-1) \cdots \zeta(s - d + 1) \, , 
\end{equation}
or in the form of a Dirichlet convolution
\begin{equation}
  f^{L_d} = \un * \N * \N^2 * \dots * \N^{d-1} \, ,  
\end{equation}
where $\N^k(n) = \set{1^k, 2^k, 3^k, \dots}$.

\end{document}